\documentclass[journal,twoside,onecolumn,draftclsnofoot]{IEEEtran}

\usepackage{amsmath,amssymb,bm} 
\usepackage{graphicx}           
\usepackage{tabularx}           
\usepackage{cite}               

\usepackage[normalem]{ulem}     
\usepackage{enumitem}           
\usepackage{etaremune}          
\usepackage[T1]{fontenc}        
\usepackage{xcolor}

\usepackage{braket}             
\usepackage{outlines}           
\usepackage[bookmarks,bookmarksnumbered,bookmarksopen=false,hidelinks]{hyperref} 

\newcommand{\der}[2]{\frac{d#1}{d#2}}

\newcommand{\pder}[2]{\frac{\partial#1}{\partial#2}}

\newcommand{\eval}[3]{\left.#1\right\vert_{#2}^{#3}} 

\renewcommand{\(}{\left(}
\renewcommand{\)}{\right)}
\renewcommand{\[}{\left[}
\renewcommand{\]}{\right]}
\newcommand{\lb}{\left\lbrace}
\newcommand{\rb}{\right\rbrace}

\begin{document}
\title{Monolayer MoS$_2$ Strained to 1.3\% with a Microelectromechanical System}
\author{Jason~W.~Christopher,~\IEEEmembership{Member,~IEEE,}
        Mounika~Vutukuru,
        David~Lloyd,
        J.~Scott~Bunch,
        Bennett~B.~Goldberg,
        David~J.~Bishop,~\IEEEmembership{Member,~IEEE,}
        and~Anna~K.~Swan~\IEEEmembership{Senior Member,~IEEE,}
\thanks{Manuscript received Mon. Day, Year. JWC thanks the Department of Defense (DoD), Air Force Office of Scientific Research for its support through the National Defense Science and Engineering Graduate (NDSEG) Fellowship, 32 CFR 168a. DJB is supported by the Engineering Research Centers Program of the National Science Foundation under NSF Cooperative Agreement No. EEC-1647837. This work was also supported by the National Science Foundation Division of Materials Research under grant number 1411008.}
\thanks{J. W. Christopher is with the Department of Physics, Boston University, Boston, MA 02215.}
\thanks{M. Vutukuru is with the Department of Electrical and Computer Engineering, Boston University, Boston, MA 02215 USA.}%
\thanks{D. Lloyd is with the Department of Mechanical Engineering, Boston University, Boston, MA 02215 USA.}%
\thanks{J. S. Bunch is with the Department of Mechanical Engineering, Boston University, Boston, MA 02215 USA.}%
\thanks{B. B. Goldberg is with the Department of Physics, Boston University, Boston, MA 02215 USA, and also with the Department of Physics and the Searle Center for Advancing Learning and Teaching, Northwestern University, Evanston, IL 60208 USA.}%
\thanks{D. J. Bishop is with the Department of Electrical and Computer Engineering and Department of Physics, Boston University, Boston, MA 02215 USA.}
\thanks{A. K. Swan is with the Department of Electrical and Computer Engineering, Boston University, Boston, MA 02215 USA e-mail: swan@bu.edu}
\thanks{This paper has supplementary downloadable material available at http://ieeexplore.ieee.org, provided by the authors. This includes discussion of strain achieved in prior work combining 2D materials with MEMS, displacement versus power characterization of our devices, and derivation of substrate dependent effective Poisson's ratio.}}

\markboth{Journal of Microelectromechanical Systems,~Vol.~VV, No.~N,~Month~YYYY}{Christopher \MakeLowercase{\textit{et al.}}: Monolayer MoS$_2$ Strained to 1.3\% with a Microelectromechanical System}

\IEEEpubid{0000--0000/00\$00.00 ~\copyright ~2018 IEEE}
\maketitle
\IEEEpeerreviewmaketitle

\begin{abstract}
We report on a modified transfer technique for atomically thin materials integrated onto microelectromechanical systems (MEMS) for studying strain physics and creating strain-based devices. Our method tolerates the non-planar structures and fragility of MEMS, while still providing precise positioning and crack free transfer of flakes. Further, our method used the transfer polymer to anchor the 2D crystal to the MEMS, which reduces the fabrication time, increases the yield, and allowed us to exploit the strong mechanical coupling between 2D crystal and polymer to strain the atomically thin system. We successfully strained single atomic layers of molybdenum disulfide (MoS$_2$) with MEMS devices for the first time and achieved greater than 1.3\% strain, marking a major milestone for incorporating 2D materials with MEMS We used the established strain response of MoS$_2$ Raman and Photoluminescence spectra to deduce the strain in our crystals and provide a consistency check. We found good comparison between our experiment and literature.
\end{abstract}
\begin{IEEEkeywords}
MEMS, Monolayer MoS$_2$, Strain, Raman, Photoluminescence.
\end{IEEEkeywords}

\section{Introduction}
\IEEEPARstart{T}{wo} dimensional (2D) materials can withstand an order of magnitude more strain than their bulk counterparts, which results in dramatic changes to electrical\cite{Settnes2016}, thermal\cite{Chen2014a} and optical properties\cite{Lloyd2016a,Conley2013a}. Ideally, we would be able to precisely control the strain field in these systems to study in detail the effect of strain and to create new strain-based devices. However, current techniques offer limited control over the strain field, and require bulky pressure chambers\cite{Lloyd2016a,Kitt2013a} or large beam bending equipment\cite{Rice2013a,Wang2013b} incompatible with most applications. Here we demonstrate that MEMS can be used to dynamically strain atomically thin materials, which provides a method for straining 2D materials that can be incorporated in technologically relevant devices.

\IEEEpubidadjcol

\begin{figure}[!t]
  \centering
  \includegraphics[width=3.5in]{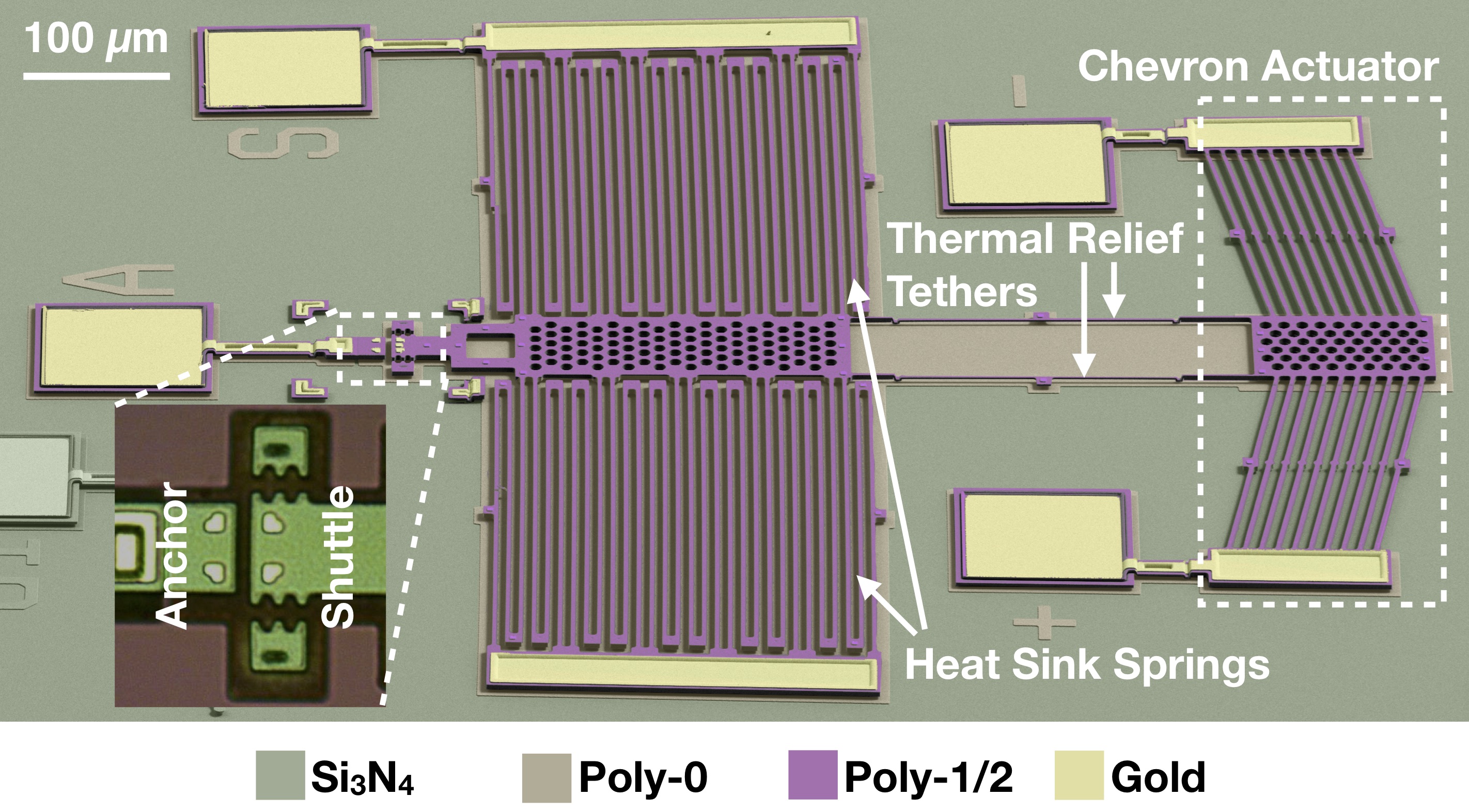}
  \caption{Colorized SEM image of a typical MEMS device used for straining 2D materials. The atomically thin crystal is placed on the left side of the MEMS across a 3 $\mu$m gap between the anchor and shuttle shown in the inset optical image. The shuttle is straddled by verniers that allow precise optical measurement of the shuttle's displacement. To the right of the sample stage there are a series of very soft springs and long thin tethers which together isolate the sample stage from the heat generated in the Chevron actuator at the far right end of the MEMS.}
  \label{fig:architecture}
\end{figure}

Previous experiments have used MEMS to strain nano-materials such as nanotubes\cite{Muoth2012a,Muoth2013}, trilayer graphene\cite{Garza2014}, and monolayer graphene \cite{Zhang2014a,Goldsche2018}. Similar to the nanotube and trilayer graphene experiments, we adopt thermally isolated chevron actuators. A colorized SEM image of one of our devices is shown in Fig.~\ref{fig:architecture}. Our devices are fabricated using MEMSCAP's PolyMUMPS\cite{polyMUMPs} process which has three poly-silicon layers. The first layer we use for grounding, and the second and third are combined to make rigid double thick structures. While 2D materials are atomically thin, it still requires significant force to strain them, as they are very stiff. Graphene, for example, is the stiffest material ever measured\cite{Lee2008a}. For this reason chevron actuators are ideal for straining our 2D crystals because they are capable of large pull forces\cite{Que1999a}. The actuator in Fig.~\ref{fig:architecture} is located on the right side. These actuators rely on thermal expansion caused by Joule heating to buckle the beams and create motion. To avoid heating the sample stage with the actuator we thermally isolate the actuator with long, thin thermal relief tethers that have a large thermal impedance. Further, we place many soft heat sink springs in parallel near the sample stage to create a low thermal impedance between the sample stage and the MEMS die. This geometry creates effectively a thermal resistor divider circuit dramatically cutting down on the heat that reaches our atomically thin samples\cite{Vutukuru2018}. The samples are placed on the stage on the left side of Fig.~\ref{fig:architecture}, shown in the inset. The stage has an anchor side that is secured to the MEMS die on the left, and a shuttle side that is connected to the actuator to stretch the 2D crystal on the right. The two sample stages are separated by a gap that is nominally 3 $\mu$m.

The most significant challenge in straining 2D materials with MEMS is anchoring the 2D crystal to the substrate. Crystals capable of forming 2D systems have strong in-plane bonds, but weak, van der Waals, out-of-plane bonds which is why it is a straight forward technique to obtain single layers from exfoliation of bulk crystals. This is also why many crystals that have become standard 2D materials are also known as good lubricants, graphene and molybdenum disulfide (MoS$_2$) for example. A number of strategies have been used to improve the bonding between atomically thin flakes and various substrates. Gold has been used with graphene in experiments stretching Kirigami structures\cite{Blees2015}, and photoresists have been used with graphene resonators to reduce damping caused by slipping between the membrane and substrate\cite{Lee2013z,Guan2015a,Goldsche2018}. However, in the majority of these examples the strain within the crystal structure is small, and hence the required anchor force is small. For example in the Kirigami structures the deformation is macroscopic and there is actually little strain within the material. Most applicable to our work is the work by Garza \emph{et al.} \cite{Garza2014} who anchored trilayer graphene using a femto-pipette to epoxy the 2D material to the MEMS. While we find this work \cite{Garza2014} to be a valuable reference for many important concepts regarding the use of MEMS to strain 2D materials, a close examination of the data shows that appreciable strain was not achieved in that experiment. For further details see supplementary material S1. We use a polymer to place the atomically thin flake onto our MEMS device. Instead of removing the polymer to obtain a clean sample, we leave the polymer coating the flake and sample stage region of the MEMS providing a strong mechanical coupling between the two. The obvious draw back of this method is that the atomically thin crystal remains coated in polymer during actuation, but this is a suitable compromise at this time to make progress in the incorporation of 2D materials with MEMS.

An additional difficulty with incorporating atomically thin crystals with MEMS are the fragile, non-planer MEMS structures. The vast majority of transfer methods in the literature are targeted at placing 2D materials on flat substrates. With MEMS we have to gracefully handle steps on the substrate that are several micrometers high. Further, we need to release our MEMS devices prior to transfer to  avoid exposing the anchoring polymer and 2D crystal to hydrofluoric acid (HF), which means that we are transferring our flakes onto very fragile structures. Not only is there the possibility of the HF degrading the transfer polymer anchoring the flakes, but there is also the possibility that the HF would etch the thin native oxide layer between the transfer polymer and poly-silicon structure greatly reducing the mechanical coupling between the two. Further, since the transfer polymer covers a large portion of the sample stage, an extended HF exposure would be necessary to release the MEMS making the native oxide etch even more likely. To circumvent these issues we have developed a technique for transferring 2D materials onto MEMS utilizing a specially designed microstructure which facilitates a gentle, non-planar compliant transfer after HF release.

In the experiments presented below we focus exclusively on monolayer MoS$_2$ for our samples. MoS$_2$ is a direct-gap semiconductor\cite{Mak2010c} with two inequivalent valleys with opposite spins\cite{Xu2014}. These properties make MoS$_2$ an interesting material for building nano-electronic devices such as transistors\cite{Radisavljevic2011a} and phototransistors\cite{Yin2012a}. The two valleys can be coherently optically addressed\cite{Mak2012,Mak2014a,Zeng2012a}, which may be useful in novel applications such as valleytronics\cite{Mak2014a} and spintronics\cite{Klinovaja2013}. For these reasons there is an extensive body of research on MoS$_2$ including the strain response of the Raman and photoluminescence (PL) spectra\cite{Conley2013a,Wang2013b,Rice2013a,Lloyd2016a}. We will rely on this literature to determine the strain in our 2D crystals from the Raman and PL spectra we measure, proving that we are able to strain atomically thin flakes with MEMS. 

\begin{figure}[!t]
  \centering
  \includegraphics[width=3.5in]{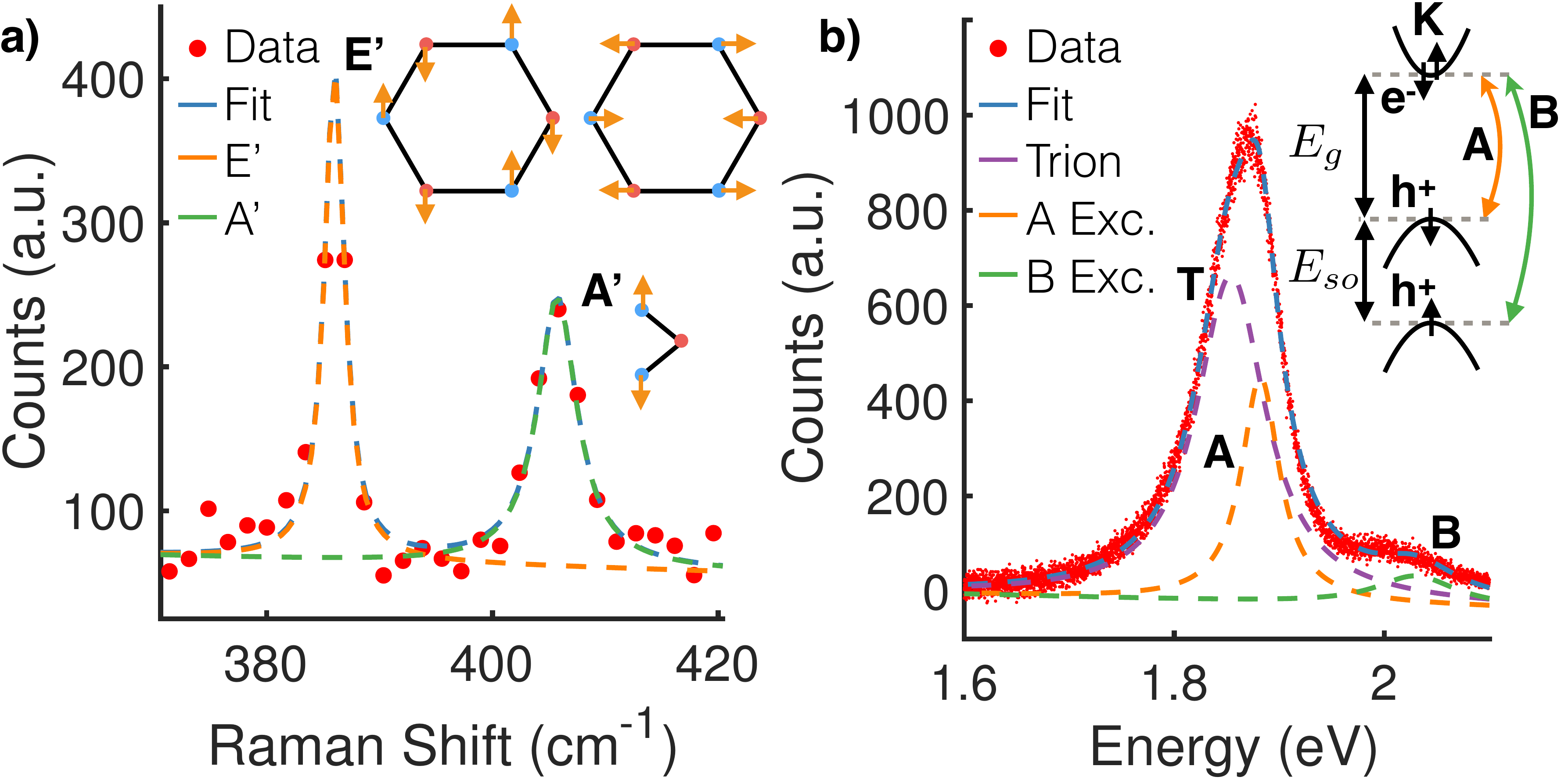}
  \caption{\textbf{a)} Raman spectrum: Next to each peak in the spectrum is a diagram depicting the motions of the atoms in the corresponding phonon mode. The red and blue circles are molybdenum and sulfur atoms. \textbf{b)} PL spectrum: The trion, A exciton, and B exciton components are broken out individually. Inset is a diagram of the band structure near the $K$ point of the BZ, which shows the spin-orbit splitting of the valence band responsible for the separation between the A and B excitons.}
  \label{fig:spectra}
\end{figure}

Fig.~\ref{fig:spectra} shows typical unstrained Raman and PL spectra for monolayer MoS$_2$. Next to each peak in the Raman spectrum in Fig.~\ref{fig:spectra}a is a diagram depicting the atomic displacements of the corresponding phonon mode. Of interest to us are the degenerate in-plane $E'$ modes with an unstrained energy of 385 cm$^{-1}$, and the out-of-plane $A'$ mode with an unstrained energy of 405 cm$^{-1}$\cite{Rice2013a,Wang2013b,Lee2010a}. Corresponding with the honeycomb crystal lattice, MoS$_2$ has a hexagonal Brillouin Zone (BZ), and like graphene the low energy electronic states occur at the $K$ and $K'$ points in the corners of the BZ where the band gap is at its minimum. The band structure near the $K$ point is shown in the inset of Fig.~\ref{fig:spectra}b, and the structure is identical at the $K'$ point but with spins flipped due to time-reversal symmetry\cite{Xu2014}. Notably the valence band is split by spin-orbit coupling which results in two exciton peaks in the PL spectrum\cite{Mak2010c}. The A and B exciton peaks correspond with the upper and lower valence bands and have unstrained energies of 1.89 eV and 2.03 eV. A third peak makes a considerable contribution to the PL spectrum and corresponds with a trion, a bound state of two electrons and hole\cite{Mak2013a,Christopher2017}. The contributions for each of these components of the PL are shown in Fig.~\ref{fig:spectra}.

\section{Methods}
\begin{figure}[!t]
  \centering
  \includegraphics[width=3.5in]{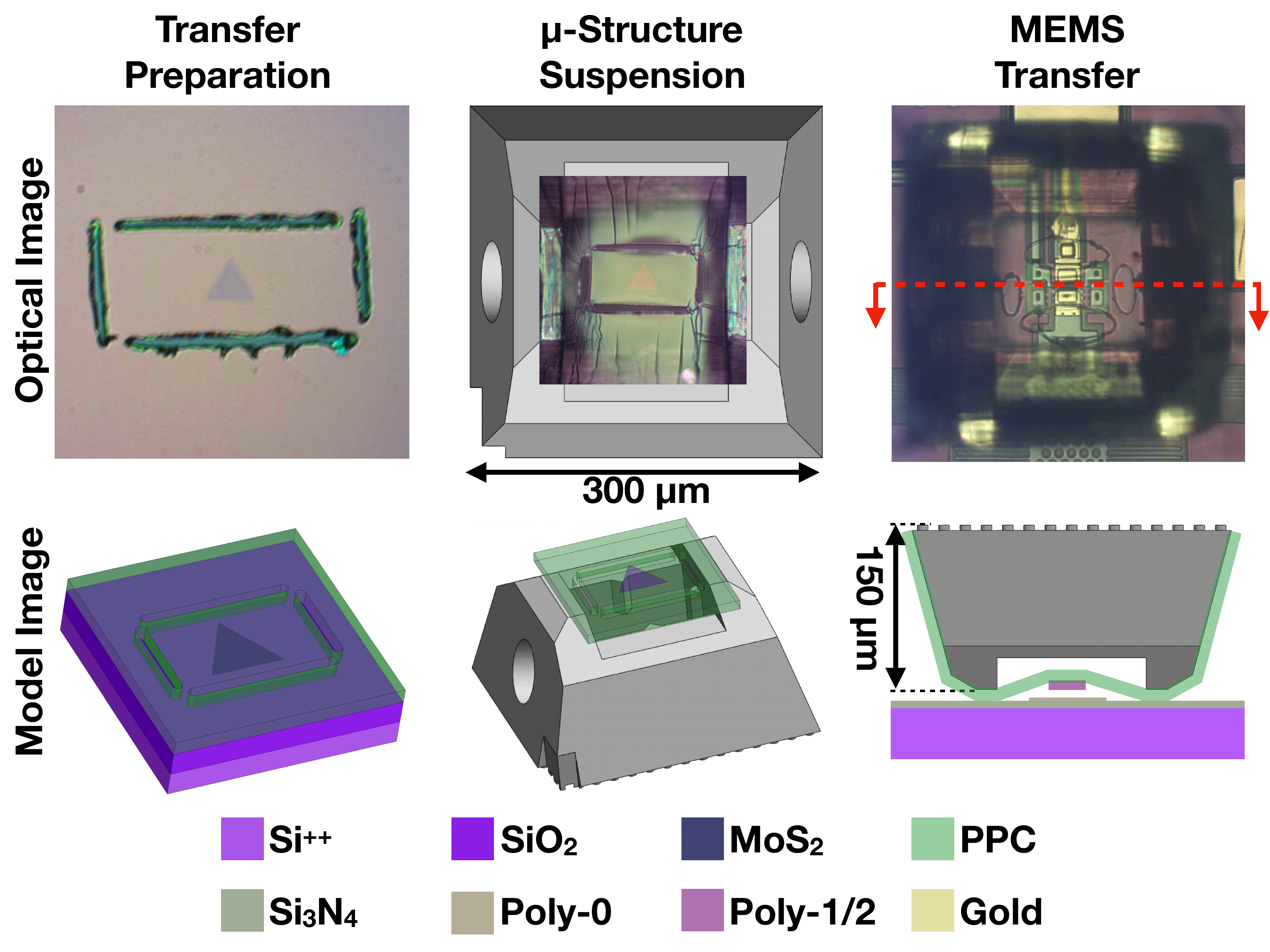}
  \caption{The top row contains optical images of actual samples and devices, while the bottom row contains model images that provide a schematic view of the fabrication method. Each column corresponds with a step in the fabrication process, and all steps except the CVD growth of the MoS$_2$ are displayed here. The microstructure shown in the last two columns has a vertical hole running through it, which is how we see the MEMS through the structure in the last column.}
  \label{fig:fabrication}
\end{figure}

\subsection{Device Fabrication}
An overview of our device fabrication method is shown in Fig.~\ref{fig:fabrication}, and has 4 steps (3 shown in the figure): sample growth, transfer preparation, microstructure suspension, and MEMS transfer. \textbf{Sample Growth} Our monolayer MoS$_2$ crystals are grown via Chemical Vapor Deposition (CVD); the details of which are provided in a previous publication\cite{Lloyd2016a} which demonstrates the high quality and strength of our MoS$_2$ films and characterizes the Raman and PL strain response using pressurized micro-chambers. Importantly the growth is done on a degenerately doped silicon substrate with 285 nm of oxide, which allows us to identify isolated monolayer flakes without cracks of suitable size, $\sim$60 $\mu$m on a side. 

\begin{figure}[!t]
  \centering
  \includegraphics[width=3.5in]{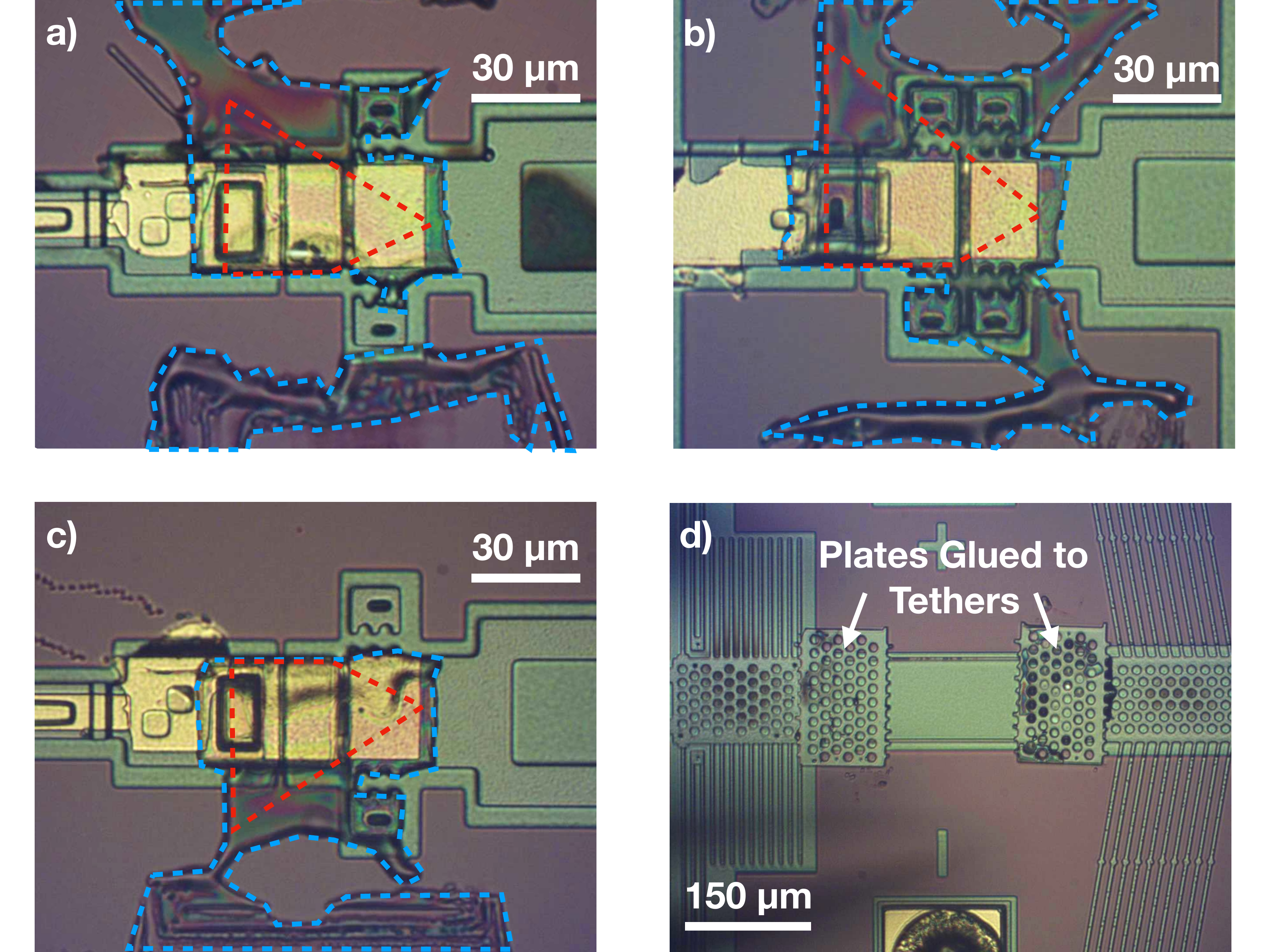}
  \caption{Optical images of three devices which showed strain response. The blue dashed line outlines the regions covered in PPC, and the red dashed line outlines the MoS$_2$ flakes on the devices. \textbf{a)} Device M24 \textbf{b)} Device M25 (Force Meter) \textbf{c)} Device M26 \textbf{d)} Shuttles from other devices glued onto the thermal relief tethers of M26 repairing it.}
  \label{fig:devices}
\end{figure}

\textbf{Transfer Preparation} Our method begins like most methods for transferring atomically thin flakes\cite{Li2009z,Suk2011,Song2013b,Wang2013e,Zomer2014}; we spin a transfer polymer onto the MoS$_2$ film which enables us to pull the flake free of its substrate and move it onto a new substrate. In our case we use Poly(propylene carbonate) (PPC), because we find it to be less brittle than the more commonly used Poly(methyl methacrylate) (PMMA). Next we deviate from the standard methods in two ways: 1) We make ``strain relief'' cuts in the transfer polymer using a probe in a micro-manipulator. These cuts allow us in the next step to tightly stretch the polymer over a microstructure without creating cracks in the sample. 2) We are able to release the transfer polymer and flake from the growth substrate with a simple deionized water bath. Notably this is safer and cleaner than typical methods that use hydrofluoric (HF) acid or other chemicals to etch the substrate away from the flake.

\textbf{Microstructure Suspension} Now that the MoS$_2$ is freely suspended on PPC we can place it on a microstructure designed specifically for use with our MEMS. The microstructure is fabricated via Direct Laser Writing (DLW), a high-resolution (sub-micrometer) 3D printing technique. The transfer of the flake and PPC onto the microstructure is accomplished using a micromanipulator for positioning, and temperature controlled stage to slightly heat the microstructure ( $\sim$35$^{\circ}$C). The heat improves adhesion and reduces strain on the polymer and flake. The optical image in Fig.~\ref{fig:fabrication} clearly shows cracks in the PPC, but the cracks stop at the strain relief cuts leaving the sample pristine.

\textbf{MEMS Transfer} 
Finally, the flake and PPC are transfered onto the MEMS. Prior to transfer the MEMS is released in HF, functionality is tested, and a thorough cleaning via oxygen plasma is done. Similar to the transfer onto the microstructure a micromanipulator is used to position the sample over the MEMS, and the MEMS is heated. This time the stage is heated to $\sim$75$^{\circ}$C in order to heat the PPC through its glass transition temperature ensuring that the PPC melts onto the MEMS structure. The temperature controlled stage also allows us to achieving a gentle transfer by using the thermal expansion of the stage to bring the sample in contact with the MEMS. The last step is to heat the MEMS to 90$^{\circ}$C on a hotplate for 10 minutes. This allows the PPC to fully melt enabling the 2D material to fully conform to the MEMS substrate. 

\begin{figure*}[!t]
  \centering
  \includegraphics[width=6in]{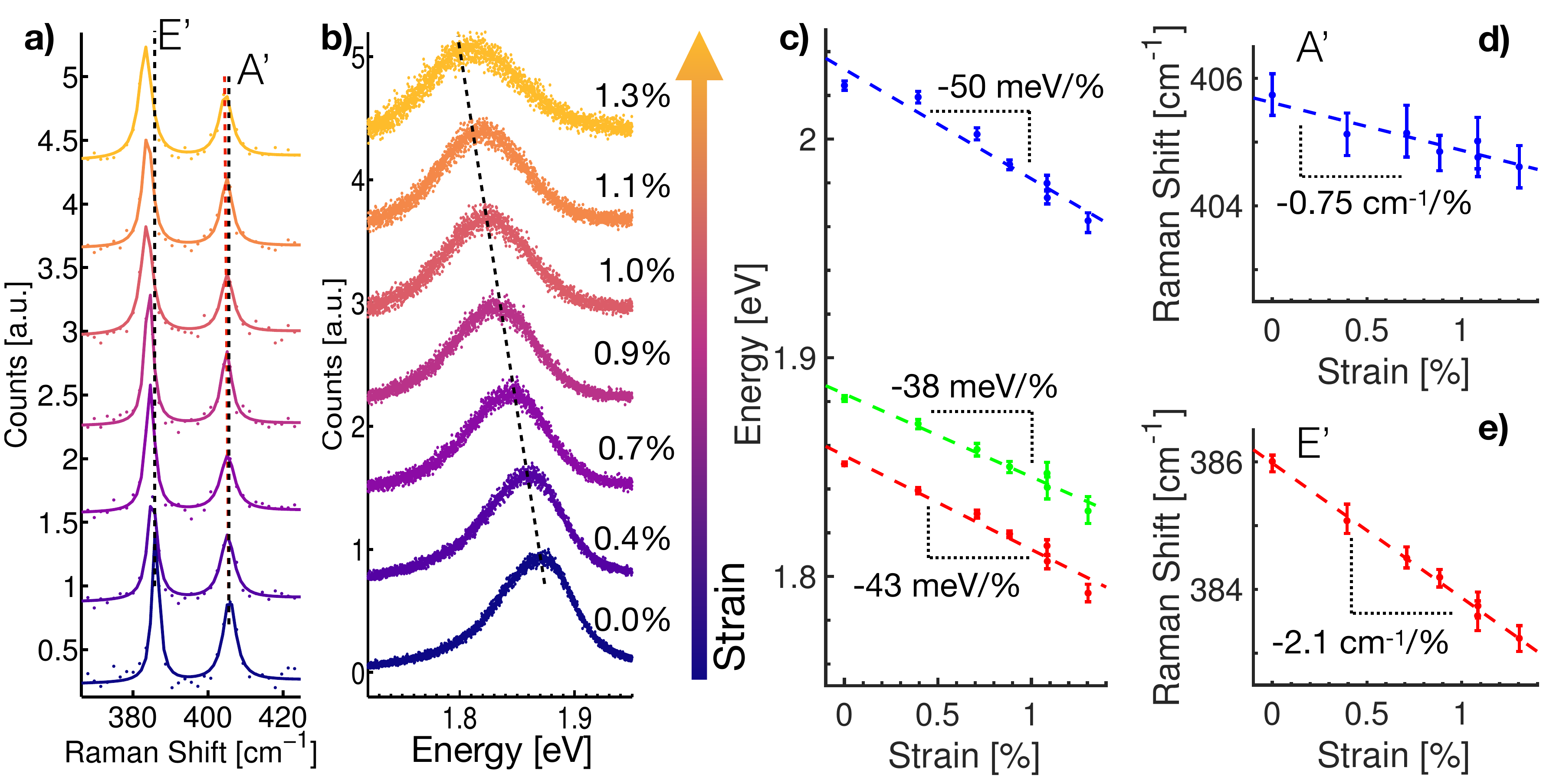}
  \caption{Raman and PL spectra of sample M26 v2 as a function of strain. Values for the strain are extracted from the peak positions as described in the text. \textbf{a)} Raman Spectra with increasing strain, dashed line indicates the unstrained peak positions.  \textbf{b)} PL Spectra, dashed line provides a guide to the eye for how the peak shifts with strain. \textbf{c)} Trion peak position versus strain in red, A exciton peak position versus strain in green, and B exciton peak position versus strain in blue. \textbf{d)} $A'$ phonon energy versus strain \textbf{e)} $E'_-$ phonon energy versus strain.}
  \label{fig:strainWaterfall}
\end{figure*} 

Fig.~\ref{fig:devices}a-c show optical images of three devices we fabricated using our method. We will refer to these devices and the datasets collected from them as M24, M25, and M26 throughout the text. Note that M25 has a different sample anchor stage from M24 and M26. The anchor on M25 is mounted on springs to measure the stress in the 2D crystal. Common to all of these devices is a coating of PPC, which has been outlined in blue in the optical images. While the PPC does provide a strong mechanical coupling between the flake and the MEMS, it is a viscoelastic polymer, which damps the actuation of the sample stage shuttle. This introduces some subtleties in interpreting our data as discussed below. From these three devices we collected four datasets. The fourth dataset comes from a second experiment with device M26 collected after repairing the device. The first experiment on M26 ended prematurely when the  thermal relief tethers fractured. We repaired the breaks in the tethers by UV gluing scavenged MEMS parts over the breaks in the tethers as shown in Fig.~\ref{fig:devices}d. We will refer to the fourth dataset, collected after repairing M26, as M26 v2. 

\subsection{Measurements}
Fig.~\ref{fig:strainWaterfall} shows an example data set consisting of Raman and PL spectra measurements made on M26 v2 under actuation. Included in Fig.~\ref{fig:strainWaterfall} are the various peak positions versus strain and the corresponding slopes giving the rate at which strain changes the peak position. These slopes are tabulated below for each data set in Table~\ref{tbl:ramanSlope} and Table~\ref{tbl:plSlope} for Raman and PL data respectively.

The Raman and PL measurements were made using a Renishaw spectrometer with an 1800 line per mm grating. The MoS$_2$ films were excited with an Argon ion laser with wavelength 514.5 nm with a beam waist of $\sim$1 $\mu$m. Considering the fragility and low thermal dissipation in our suspended samples, power was kept below 20 $\mu$W to avoid heating and sample damage. Heat from the Chevron actuator can significantly elevate the temperature of the entire die if a good path for thermal dissipation is not established. To prevent this our dies are mounted with silver epoxy to a copper plate which can be cooled with a thermal electric cooler (TEC) as needed. The copper plate has an internal platinum resistive thermal device for monitoring the temperature, and we periodically check the temperature of the sample anchor and shuttle stages using Raman thermometry\cite{Garza2014}. For our actuators we are able to sufficiently strain the samples with less than 250 mW of power, and we find that as long as the die has a low thermal impedance to the copper plate, the TEC is not necessary.

Since friction between the MoS$_2$ film and MEMS is low, it is important that we are able to detect strain at its earliest onset. We achieve this by continuously monitoring the A exciton peak, the strongest peak, while slowly increasing the power to the actuator in steps which will increase the strain by less than 0.5\%. The step size is determined by a precise measurement of the size of the suspended portion of the sample, always underestimating for safety, and relying on our calibrated displacement versus power curve for our devices (see supporting material S2). As soon as a shift is noticed in the peak position, the power is held constant while taking PL and Raman measurements. Because of the viscoelastic behavior of our polymer anchor, in most cases it was possible to simply wait $\sim$10 minutes between measurements for the peak positions to shift further. The wait-measure cycle was repeated until the peak either stopped shifting, in which case more power was applied, or the peak relaxed marking either a major slip between the PPC and MEMS or in several cases the thermal relief tethers breaking. While in some cases it was possible to get further strain response from the sample, all datasets we have analyzed here are monotonic in A exciton peak shift. 

Given the viscoelastic behavior of our devices the amount of time spent measuring spectra becomes an important trade-off between collecting high quality data (long time) and strain resolution (short time). We found that at a minimum, we needed to collect Raman data for 135 s to have adequate statistics for our analysis. The PL has a much stronger signal, only requiring 10 s, but because of the large spectral range of the measurement the grating must be rotated during the acquisition so the measurement takes $\sim$3 min. A lower resolution grating could not be used to shorten this time, since the Raman features are narrow, only $\sim$10 data points per peak. The measurement time results in a small time delay between the measurement of each of the PL peaks, and an even larger time delay between when the Raman and PL data are collected. The implications of these time delays on our data is discussed in the sections below. 

\section{Determining Strain from Peak Positions}
Strain changes the electron and phonon band structures, which we measure as shifts in the Raman and PL peak positions. Group theory places strong restrictions on the functional dependence of peak positions on the strain tensor. Further, since the strain is small, we limit our analysis to first order in strain. For our purposes there are only two point group representations of the crystal symmetry that are of interest, $A'$ and $E'$. $A'$ is the trivial representation and must be rotationally invariant. The only first order rotational invariant of the strain tensor is the trace, $\epsilon_{xx}+\epsilon_{yy}$, which is also called the hydrostatic strain since it is the strain that is experienced when a material is compressed on all sides equally as is the case when compressing with a fluid. Hence peaks that transform under the $A'$ representation must change under uniaxial strain according to the formula
  \begin{align}
    \omega_{A'} & = \omega_{0A'}\[1 - \gamma_{A'}\(1-\nu\)\epsilon\] \label{eq:Astrain}
  \end{align}

where $\omega_{0A'}$ is the zero strain energy, $\gamma_{A'}$ is the Gr\"uneisen parameter, $\nu$ is the Poisson's ratio, and $\epsilon$ is the magnitude of the uniaxial strain. Both the Raman $A'$ peak and the PL peaks shift under strain according to~(\ref{eq:Astrain}), and we'll use the above notation for the $A'$ phonon and replace $\omega$ with $E$ and $A'$ with $A$ when referring to the A exciton peak.

The $E'$ phonon peak is a degenerate peak that shifts under uniaxial strain according to
   \begin{align}
    \omega_{E'}^{\pm} & = \omega_{0E'}\lb1-\[\gamma_{E'}\(1-\nu\)\mp\frac{\beta_{E'}}{2}\(1+\nu\)\]\epsilon\rb \label{eq:Estrain}
  \end{align}
where $\omega_{0E'}$ is the zero strain energy, $\gamma_{E'}$ is the Gr\"uneisen parameter (this term is identical with strain term in~(\ref{eq:Astrain})), and $\beta_{E'}$ is the shear deformation potential. The $\pm$ in~(\ref{eq:Estrain}) denotes the lifting of the degeneracy under strain, which splits the peak into two peaks, the $+$ peak and the $-$ peak.

\begin{table}[!t]
  \renewcommand{\arraystretch}{1.3}
  \renewcommand{\tabcolsep}{4pt}
  \caption{Gr\"uneisen, $\gamma$, and shear deformation potentials, $\beta$, for Raman and PL peaks of MoS$_2$}
  \label{tbl:peakParams}
  \centering
    \begin{tabular}{c c c c c}
      \hline\hline
      Reference & $\gamma_{E'}$ & $\beta_{E'}$ & $\gamma_{A'}$ & $\gamma_{A}$ \\
      \hline
      Lloyd \textit{et al.}\cite{Lloyd2016a} & 0.68 $\pm$ 0.1 & NA & 0.21 $\pm$ 0.1 & 2.6 $\pm$ 0.2\\
      Rice \textit{et al.}\cite{Rice2013a} & 0.65 $\pm$ 0.1 & 0.34 $\pm$ 0.1 & 0.21 $\pm$ 0.1 & NA \\
      Wang \textit{et al.}\cite{Wang2013b} & 0.6 $\pm$ 0.1 & 0.3 $\pm$ 0.1 & NA & NA \\
      Conley \textit{et al.}\cite{Conley2013a} & 1.1 $\pm$ 0.2\textsuperscript{*} & 0.68 $\pm$ 0.1\textsuperscript{*} & NA & 3.7 $\pm$ 0.6 \\
      \hline
      Effective Values\textsuperscript{a} & 0.64 $\pm$ 0.06 & 0.32 $\pm$ 0.07 & 0.21 $\pm$ 0.07 & 2.7 $\pm$ 0.2 \\
      \hline\hline
    \end{tabular}
    \begin{flushleft}
    \small\textsuperscript{a}Effective values and standard deviations are the maximum likelihood values and distribution standard deviation assuming the corresponding Gaussian distributions for the literature values.
    \small\textsuperscript{*}Values not used in computing effective values because of substantial disagreement with the rest of the literature.
    \end{flushleft}
\end{table}

Table~\ref{tbl:peakParams} provides a list of measured Gr\"uneisen parameters and shear deformation potential values for the various Raman and PL peaks of MoS$_2$. References that did not provide a measure of a certain parameter have Not Available (NA) listed in the table. Most experiments reported errors only for their measurement of the shift rate of the peak position with respect to strain, and not for the values of Gr\"uneisen parameter or shear deformation potential. So most errors reported in Table~\ref{tbl:peakParams} are adapted from the shift rate errors in the literature. The notable exception to this is \cite{Conley2013a}, which reported error bars for $\gamma_{E'}$ and $\beta_{E'}$. However, these parameter values disagree substantially from the rest of the literature, which is consistent, so they have not been used in computing the effective values at the bottom of the table.

Our objective is to use the known formulas for the strain behavior of the peaks, along with the parameters in Table~\ref{tbl:peakParams} to infer the strain in our measurements from the observed peak positions. However, to do so a value for the Poisson's ratio, $\nu$, must be provided. It is generally assumed that an atomically thin flake will inherit the Poisson's ratio of its substrate since it is assumed the two stick to each other well. We'll address this assumption more directly below, but for the time being adopt this assumption. The Poisson's ratio of PPC is not known, so is approximated from two similar polymers, Poly(bisphenol A carbonate) with $\nu$ = 0.41 and Polypropylene with $\nu$ = 0.43~\cite{polymer2003}. Thus in the analysis that follows $\nu$ is assumed to have a value of 0.42. For comparison, $\nu\approx$ 0.27 for monolayer MoS$_2$~\cite{Liu2014,Cooper2013a,Bertolazzi2011a}. 

\begin{figure}[!t]
  \centering
  \includegraphics[width=3.5in]{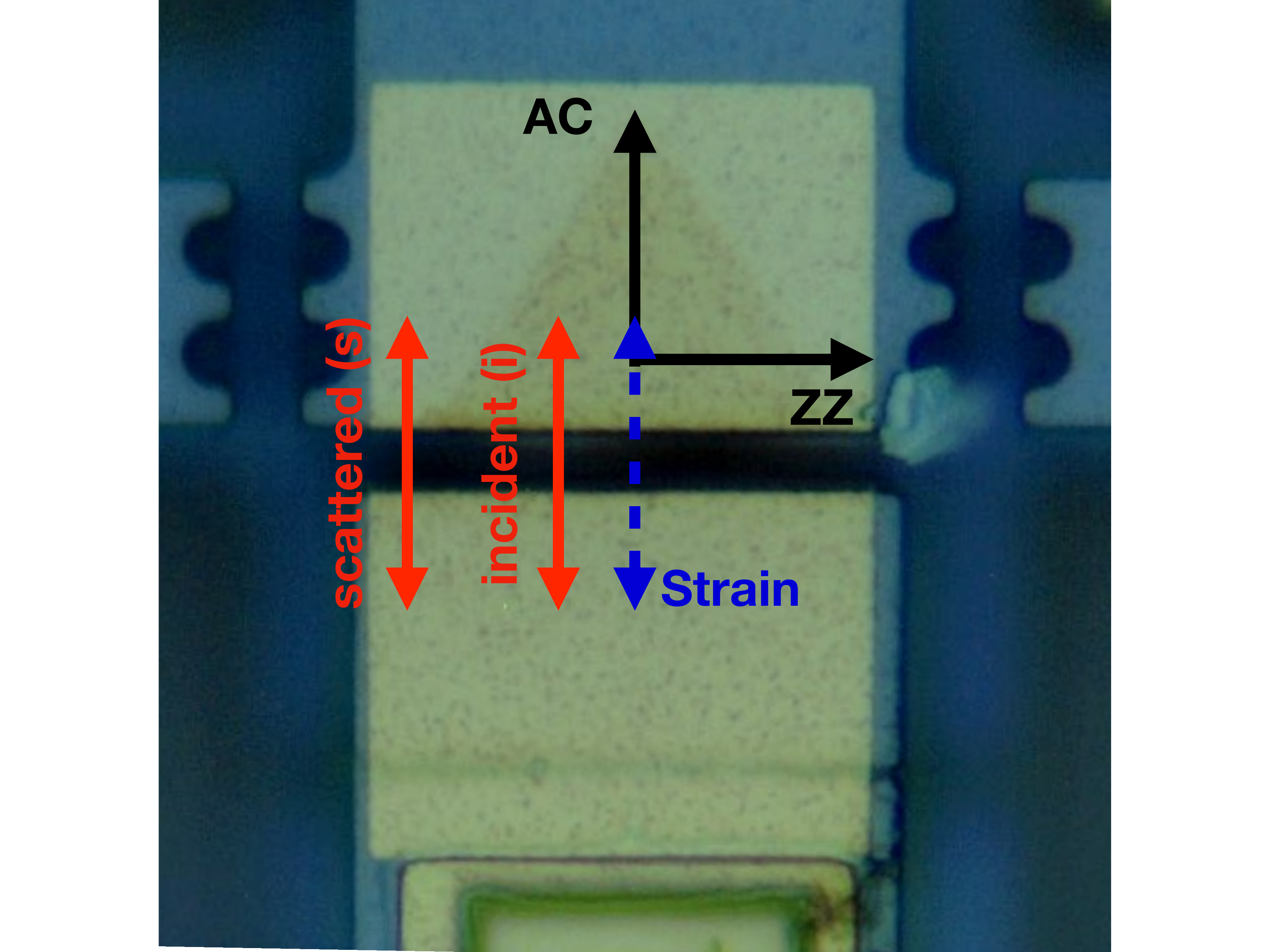}
  \caption{Orientation of crystal and incident and scattered polarizations that lead to accidental selection of only the $E'_-$ mode.}
  \label{fig:ramanSelection}
\end{figure}

An additional consideration that needs to be made in analyzing the data regards the degeneracy of the $E'$ mode which is lifted under uniaxial strain. As the strain breaks the crystal symmetry, the two degenerate $E'$ modes split into a mode that is parallel with the major strain axis, $E'_-$, and a mode that is perpendicular to the major strain axis, $E'_+$. However, the Raman spectra in Fig.~\ref{fig:strainWaterfall}a does not show split $E'$ modes. This is due to the accidental selection of only the $E'_-$ mode in our measurement setup. The selection rules for the two modes are
\begin{align}
  I_- & \propto\sin^2\(\theta_i+\theta_s+3\phi_{\epsilon}\) \\
  I_+ & \propto\cos^2\(\theta_i+\theta_s+3\phi_{\epsilon}\)
\end{align}
where $\phi_{\epsilon}$ is the angle between the ZZ axis of the crystal lattice and the major strain axis, and $\theta_i$ and $\theta_s$ are the incident and scattered polarizations of light relative to the major strain axis in the Raman measurement\cite{Rice2013a,Huang2009a,Wang2013b,Lee2017}. Fig.~\ref{fig:ramanSelection} shows an image of one of our MoS$_2$ films after the PPC has been removed making the crystal orientation obvious, and includes markers showing the incident and scattered polarizations of light selected in our experiment. It is highly preferential for CVD MoS$_2$ to grow with ZZ edge termination\cite{VanderZande2013a,Lauritsen2007,Byskov2000}. Since the flakes are placed on the MEMS pointing along the direction of strain, $\phi_{\epsilon}=90^{\circ}$. The accidental scattered polarization selection results from the relative transmission efficiency of the grating in our spectrometer, which transmits the vertical polarization with 10$\times$ the efficiency of the horizontal polarization, hence $\phi_s=0$. The laser has vertical polarization so $\phi_i=0$ as well. This combination of angles makes $I_+\approx 0$.

To extract the strain from the Raman and PL spectra, the spectra are first fit to determine the energy position of the A exciton, $E'$, and $A'$ peaks. The PL and Raman peak positions are used individually to calculate a value of strain. Then these individual strain values are combined, accounting for relative confidence intervals using standard statistical methods. An example of this analysis is shown in Table~\ref{tbl:strains} for the M26 v2 data set. The PL are fit with three Lorentzian peaks (one each for the trion, A exciton and B exciton) and a linear background.  The Raman spectra are fit with two Lorentzian peaks (one each for the $E'$ and $A'$ modes) and a linear background. We find that the maximum strain achieved in our experiments before flake slipping or device breaking occur is 1.3 $\pm$ 0.1\%. Table~\ref{tbl:maxStrain} shows the maximum change in strain observed and the pre-strain in each device. Pre-strain is calculated assuming the unstrained A exciton energy is 1.88 eV, the $E'$ phonon energy is 386 cm$^{-1}$, and $A'$ phonon energy is 405.5 cm$^{-1}$. Table~\ref{tbl:ramanSlope} and Table~\ref{tbl:plSlope} show the slopes for the Raman and PL peaks respectively for each dataset and the expected slopes given the literature values for the Gr\"uneisen and shear deformation potential for each peak.

\begin{table}[!t]
  \renewcommand{\arraystretch}{1.3}
  \caption{Composite strain analysis for M26 v2 data set}
  \label{tbl:strains}
  \centering
    \begin{tabular}{c | c c c | c}
      \hline\hline
      Step & A Exciton [\%] & $A'$ Phonon [\%] & $E'$ Phonon [\%] & Composite [\%]\\
      \hline
      0 & -0.04 $\pm$ 0.04 & -0.49 $\pm$ 0.66 & 0.00 $\pm$ 0.06 & -0.03 $\pm$ 0.03 \\
      1 & 0.34 $\pm$ 0.07 & 0.76 $\pm$ 0.68 & 0.40 $\pm$ 0.10 & 0.37 $\pm$ 0.06\\
      2 & 0.74 $\pm$ 0.10 & 0.72 $\pm$ 0.82 & 0.65 $\pm$ 0.07 & 0.68 $\pm$ 0.06\\
      3 & 1.01 $\pm$ 0.09 & 1.31 $\pm$ 0.56 & 0.78 $\pm$ 0.06 & 0.86 $\pm$ 0.05\\
      4 & 1.11 $\pm$ 0.19 & 0.98 $\pm$ 0.81 & 1.04 $\pm$ 0.10 & 1.06 $\pm$ 0.09\\
      5 & 1.33 $\pm$ 0.17 & 1.49 $\pm$ 0.58 & 0.97 $\pm$ 0.09 & 1.06 $\pm$ 0.08\\
      6 & 1.69 $\pm$ 0.21 & 1.79 $\pm$ 0.67 & 1.19 $\pm$ 0.09 & 1.28 $\pm$ 0.08\\
      \hline\hline
    \end{tabular}
\end{table}

\begin{table}[!t]
  \renewcommand{\arraystretch}{1.3}
  \caption{Maximum change in strain achieved and pre-strain}
  \label{tbl:maxStrain}
  \centering
    \begin{tabular}{c c c}
      \hline\hline
      Device & Max. Change in Strain [\%] & Pre-Strain [\%] \\
      \hline
      M24 & 0.76 $\pm$ 0.08 & -0.01 $\pm$ 0.09 \\
      M25 & 0.63 $\pm$ 0.05 & -0.05 $\pm$ 0.03 \\
      M26 & 0.86 $\pm$ 0.11 & 0.03 $\pm$ 0.06 \\
      M26 v2 & 1.30 $\pm$ 0.09 & -0.03 $\pm$ 0.03 \\
      \hline\hline
    \end{tabular}
\end{table}

\begin{table}[!t]
  \renewcommand{\arraystretch}{1.3}
  \caption{Slopes for Raman peak strain response}
  \label{tbl:ramanSlope}
  \centering
    \begin{tabular}{c c c}
      \hline\hline
      Device & $E'_{-}$ [cm$^{-1}$/\%] & $A'$ [cm$^{-1}$/\%] \\
      \hline
      M24 & -1.95 $\pm$ 0.12 & -0.78 $\pm$ 0.68 \\
      M25 & -2.28 $\pm$ 0.43 & -0.73 $\pm$ 0.52 \\
      M26 & -1.45 $\pm$ 0.49 & -0.75 $\pm$ 0.57 \\
      M26 v2 & -2.11 $\pm$ 0.19 & -0.75 $\pm$ 0.46 \\
      \hline
      Lit. & -2.32 $\pm$ 0.41 & -0.49 $\pm$ 0.17\\
      \hline\hline
    \end{tabular}
\end{table}

\begin{table}[!t]
  \renewcommand{\arraystretch}{1.3}
  \caption{Slopes for PL peak strain response}
  \label{tbl:plSlope}
  \centering
    \begin{tabular}{c c c c c}
      \hline\hline
      Device & Trion [meV/\%] & A Ex. [meV/\%] & B Ex. [meV/\%] \\
      \hline
      M24 & -55 $\pm$ 5 & -45 $\pm$ 9 & -71 $\pm$ 13 \\
      M25 & -84 $\pm$ 12 & -81 $\pm$ 20 & -29 $\pm$ 14 \\
      M26 & -110 $\pm$ 22 & -72 $\pm$ 25 & -74 $\pm$ 21 \\
      M26 v2 & -43 $\pm$ 2 & -38 $\pm$ 3 & -50 $\pm$ 5 \\
      \hline
      Lit. & - & -30 $\pm$ 2 & - \\
      \hline\hline
    \end{tabular}
\end{table}

\section{Strain Response Comparison with Literature}
Since strain in our experiments is determined from literature values of the strain response of the various Raman and PL peaks, any attempt to compare the strain response we observe with literature would be circular reasoning. However, if we take ratios of the strain response of the peaks, then we eliminate strain as an independent variable, and create truly independent measures that can be compared with literature. Since there are three peaks with known strain response we can create three ratios of strain responses:
\begin{align}
  \der{\omega_{A'}}{E_A} & = \frac{\der{\omega_{A'}}{\epsilon}}{\der{E_A}{\epsilon}} = \frac{\omega_{0A'}}{E_{0A}}\frac{\gamma_{A'}}{\gamma_A},\\
  \der{E_A}{\omega_{E'}^-} & = \frac{\der{E_A}{\epsilon}}{\der{\omega_{E'}^-}{\epsilon}} = \frac{E_{0A}}{\omega_{0E'}}\frac{\gamma_A\(1-\nu\)}{\gamma_{E'}\(1-\nu\)+\frac{\beta_{E'}}{2}\(1+\nu\)},\\
  \der{\omega_{A'}}{\omega_{E'}^-} & = \frac{\der{\omega_{A'}}{\epsilon}}{\der{\omega_{E'}^-}{\epsilon}} = \frac{\omega_{0A'}}{\omega_{0E'}}\frac{\gamma_{A'}\(1-\nu\)}{\gamma_{E'}\(1-\nu\)+\frac{\beta_{E'}}{2}\(1+\nu\)}\label{eq:dEdA}.
\end{align}
Table~\ref{tbl:expVlit} contains values for the ratios computed from the experimental data and from the Gr\"uneisen parameter and shear deformation potential values from the literature. In calculating the 1$\sigma$ confidence interval for the literature values of the ratios, it was necessary to assume log normal distributions for the parameters. This is because the uncertainty is large relative to the parameter values and the parameters must be non-negative. Further, 1$\sigma$ confidence intervals were not available for all sources. In such cases we have assumed the interval to be equal to the worst reported interval for the same parameter by an alternative source. The interval for $\gamma_{A'}$ has to be completely assumed since none of the sources provide an interval. We have assumed the confidence interval to be $\pm$ 0.1 for each measurement, the largest interval for any of the Raman parameters used in the analysis. Given the non-linear functional form of the slope ratios, non-normal distribution for the parameters, and large uncertainties, the literature values for the ratios were computed using the Monte Carlo method with 10$^7$ samples. Several calculations with a smaller number of samplings were done to ensure convergence of the calculation.

\begin{table}[!t]
  \renewcommand{\arraystretch}{1.3}
  \renewcommand{\tabcolsep}{2pt}
  \caption{Relative shift rates of peaks}
  \label{tbl:expVlit}
  \centering
    \begin{tabular}{c c c c c c}
      \hline\hline
      Slope Ratio & M24 & M25 & M26 & M26 v2 & Lit. \\
      \hline
      $\der{\omega_{A'}}{E_A}$ [cm$^{-1}$/eV] & 18 $\pm$ 6 & 10 $\pm$ 3 & 8 $\pm$ 3 & 20 $\pm$ 5 & 17 $\pm$ 22 \\
      $\der{E_A}{\omega_{E'}^-}$ [meV/cm$^{-1}$] & 23 $\pm$ 3 & 36 $\pm$ 5 & 40 $\pm$ 8 & 18 $\pm$ 1 & 15 $\pm$ 6 \\
      $\der{\omega_{A'}}{\omega_{E'}^-}$ [-] & .41 $\pm$ .15 & .32 $\pm$ .12 & .43 $\pm$ .17 & .35 $\pm$ .10 & .24 $\pm$ .34 \\
      \hline\hline
    \end{tabular}
\end{table}

\section{Discussion}
The data display clear inconsistencies. The most obvious trend is that devices M25 and M26 behave differently from M24 and M26 v2. M25 and M26 both have inconsistent slopes for the PL peaks. Not only is this different from the behavior of M24 and M26 v2, but also from experiments on biaxially strained MoS$_2$~\cite{Lloyd2016a} where all the peaks shift with roughly the same slope. This behavior is also apparent in Table~\ref{tbl:expVlit} where it results in low values for $\der{\omega_{A'}}{E_A}$ and high values for $\der{E_A}{\omega_{E'}^-}$. Further, $\der{\omega_{A'}}{E_A}$ is arguably the most important comparison we can make with literature since it is independent of the Poisson's ratio. However, the M25 and M26 values for $\der{\omega_{A'}}{E_A}$ are significantly lower than the values for M24 and M26 v2. In the case of M25 we have good reason to be suspicious of the consistency of the data because the sample stage was designed to shift in order to act as a force meter. We believe that the viscoelasticity of the anchor polymer allowed the sample stage to slowly creep, reducing the strain between the PL and Raman measurements. This hypothesis is supported by the fact that $\der{\omega_{A'}}{\omega_{E'}^-}$, a comparison between two Raman peaks, is consistent with the other data sets. As for the M26 data, we suspect that one of the tethers connecting the actuator to the sample shuttle broke while acquiring the data and that the failure only came to our attention after the second tether broke. This would have caused the strain to partially deviate from uniaxial, and could have also introduced a similar reduction in strain between PL and Raman measurements. Like the M25 data, $\der{\omega_{A'}}{\omega_{E'}^-}$ for the M26 data is consistent with the other two data sets suggesting some slip between Raman and PL measurements. Considering the uncertainty regarding the M25 and M26 data sets we will disregard them in our discussion below. 

Now we return to the peak slopes in Table~\ref{tbl:ramanSlope} and Table~\ref{tbl:plSlope}. For the M24 and M26 v2 data the slopes are consistent for the trion, A exciton and B exciton as expected from literature~\cite{Lloyd2016a}, and the discrepancy between the data sets is not much more than two standard deviations. Similarly, the literature value for the A exciton slope is less than two standard deviations from the slopes for either data set, but does appear low. Turning to the Raman slopes, we see that while the $A'$ slope is consistent across all data sets, there is a very large uncertainty in its value. The large uncertainty is do to the small shift in the $A'$ peak under strain and low amplitude of the peak, which result in low confidence in the peak position and thus slope. Though the literature value for the $A'$ slope is within a single standard deviation of all the measured slopes, our data suggest that the literature value is low. There is also good agreement between the $E'_-$ slope of the M24, M26 v2, and literature values, and that all three slope ratios in Table~\ref{tbl:expVlit} are in good agreement between M24, M26 v2 and the literature values. This gives us good confidence in the strain values we have derived from the data.

We have assumed throughout that the MoS$_2$ flakes inherit the Poisson's ratio of the PPC. Here we evaluate the validity of that assumption. In other experiments that strain 2D materials on a substrate, strain is calculated in the substrate, neglecting the small perturbations caused by the atomically thin flake. Then it is assumed that there is no slipping between the flake and the substrate, so the strain in the flake must be the same as in the substrate. Hence, the flake inherits the Poisson's ratio of the substrate. However, in our case, the 2D material is much more than a perturbation to the strain distribution. For mechanical calculations, the effective Young's modulus and thickness of MoS$_2$ are $\sim$270 GPa and 0.65 nm~\cite{Liu2014,Bertolazzi2011a}, while the Young's modulus of PPC is $\sim$37 MPa~\cite{Nechifor2009,Jiang2016a} and we estimate the thickness to be no greater than 600 nm given the optical interference of comparably prepared films of PPC on silicon substrates. Thus the effective 2D Young's modulus of MoS$_2$ and PPC are 175 N/m and 22 N/m respectively, and it is no longer a good assumption that the substrate elastic constants alone determine the strain distribution. In the supplementary material, S3, we discuss the boundary conditions and a first-order method for estimating the effective Poisson's ratio for 2D materials adhered to substrates. Importantly, the effective Poisson's ratio depends on the thickness of the PPC substrate, and this potentially explains the discrepancies between the experimental and literature values for the Raman and PL peak slopes in Table~\ref{tbl:ramanSlope} and Table~\ref{tbl:plSlope}. Further, since the PPC could be of slightly different thicknesses on the M24 and M26 v2 devices, the effective Poisson's ratios could be slightly different which would explain some of the device to device variation in peak slopes.

A final source of error worthy of discussion is the difference in strain due to viscoelastic drift between the time of the PL and Raman measurements. We did attempt to correct for the drift in strain in the M24 dataset by measuring the peak shift versus time in the Raman and PL data, and then take ratios of the shift rates to determine values of $\der{\omega_{A'}}{E_A}$, $\der{E_A}{\omega_{E'}^-}$, and $\der{\omega_{A'}}{\omega_{E'}^-}$. However, the changes to the slope ratios were much less than the confidence intervals, and there is significant uncertainty as to the exact time delay, so we chose not to include this correction in our analysis.

\section{Conclusion}
In conclusion, we have strained monolayer MoS$_2$ with a MEMS for the first time, and achieved 1.3 $\pm$ 0.1\% strain. This is a major milestone in the field of 2D materials and MEMS, and marks an important advancement towards creating novel devices with 2D materials. While there is much work to be done in improving the sample quality and anchoring of the 2D material, this opens a direct path towards building novel strain based devices such as strain tunable LEDs, FETs, and even a low resolution spectrometer by adjusting the absorption spectrum. Further, the MEMS platform offers many exciting avenues for exploring physics in 2D systems by enabling strain engineering. Some obvious examples include Pseudo-magnetic field generation and exciton confinement for forming exciton condensates.

\section*{Acknowledgment}
The Author's thank Rachael Jayne for fabricating the microstructures used for transferring the 2D materials in this paper.

\bibliographystyle{IEEEtran}
\bibliography{IEEEabrv,MEMSstrainedMoS2}

\begin{IEEEbiography}[{\includegraphics[width=1in,height=1.25in,clip,keepaspectratio]{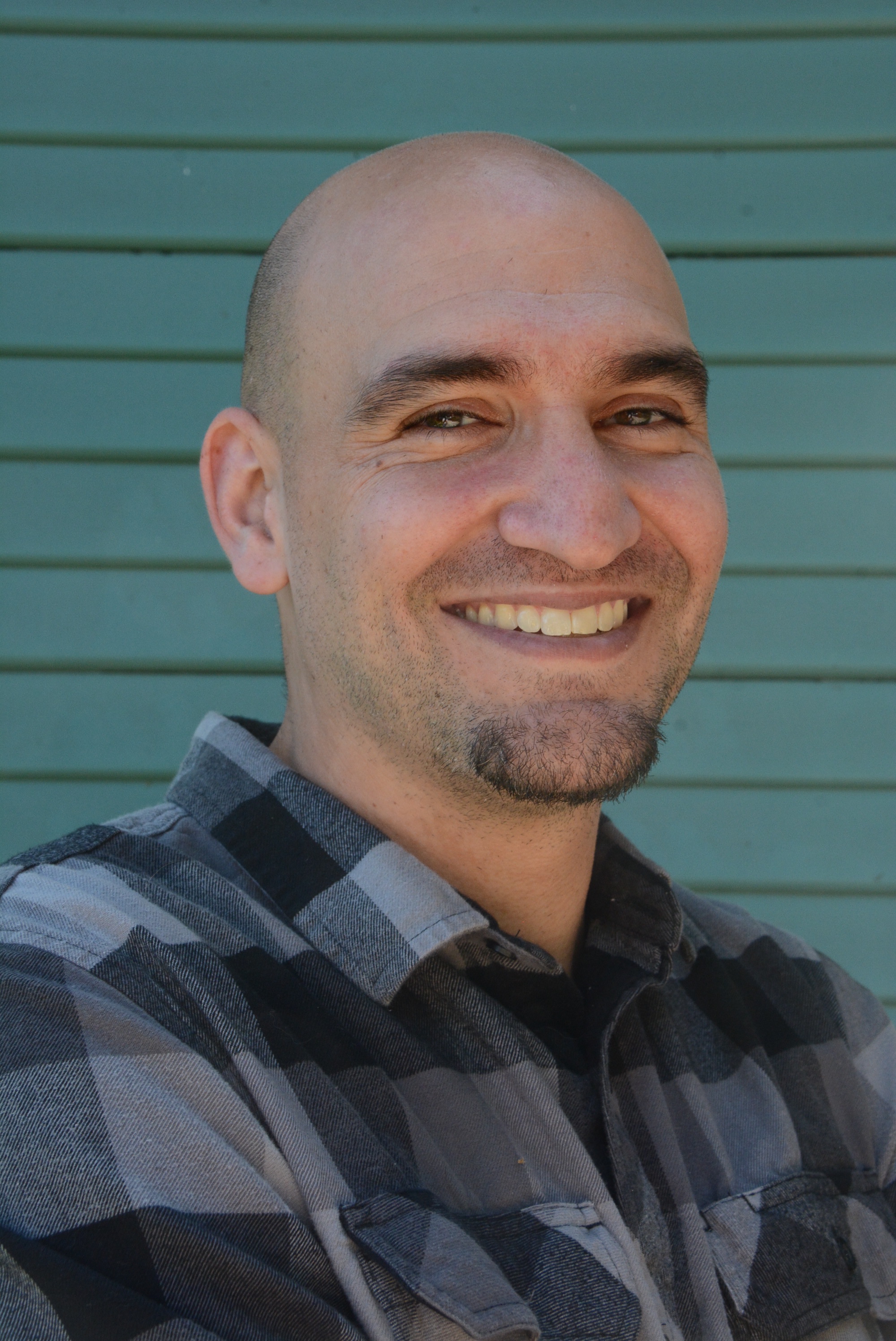}}]{Jason W. Christopher} (M'09) received the B.S. degree in Physics and Electrical Engineering and Computer Science from Massachusetts Institute of Technology in 2005, and spent several years working as an electrical engineer in Silicon Valley, most notably as a technical lead for the Trackpad Team at Apple, Inc. He then received the M.S. and Ph.D. degrees in Physics from Boston University in 2017 and 2018, respectively, where he studied mechanisms for controlling strain in 2D materials and measured the effect of strain on material properties using Raman and Photoluminescence spectroscopies. Currently he is applying his expertise in electronics, measurement, and statistical inference to develop unobtrusive methods for measuring patient vital signs and creating models to predict patient wellness at Myia Labs, Inc.
\end{IEEEbiography}

\begin{IEEEbiography}[{\includegraphics[width=1in,height=1.25in,clip,keepaspectratio]{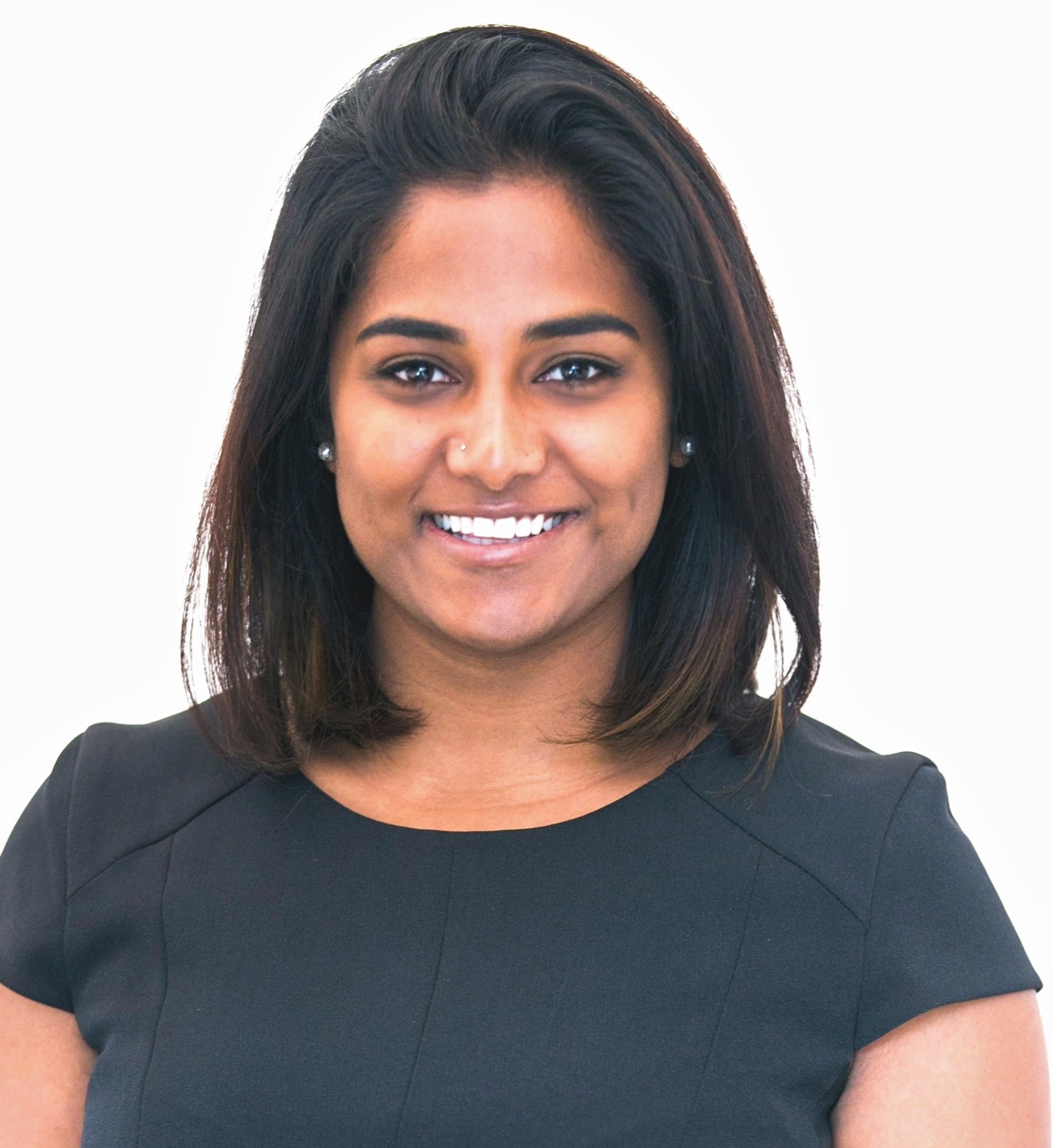}}]{Mounika Vutukuru} received the B.S. degree in electrical engineering with a concentration in Nanotechnology and the B.A. degree in physics from Boston University in 2015. She is currently working toward the Ph.D. degree in electrical engineering at Boston University. 
Her research interests include strain engineering of 2D materials using microelectromechanical systems for the purposes of probing unique physics and prototyping novel electronic devices.
\end{IEEEbiography}

\begin{IEEEbiography}[{\includegraphics[width=1in,height=1.25in,clip,keepaspectratio]{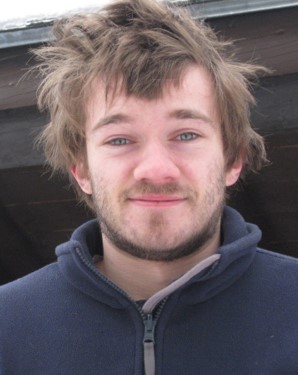}}]{David Lloyd} received an MPhys degree in physics from the University  of Oxford in 2013. He is currently studying for a Ph.D. in the department of mechanical engineering at Boston University under the supervision of Prof. Scott Bunch. He is pursuing research into the mechanical properties of 2D materials, and in particular how strain and adhesion affect atomically thin membranes, and how 2D materials can be used as separation membranes.
\end{IEEEbiography}

\begin{IEEEbiography}[{\includegraphics[width=1in,height=1.25in,clip,keepaspectratio]{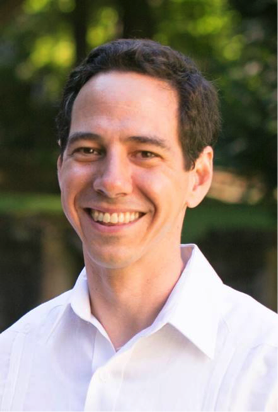}}]{J.~Scott~Bunch} is an Associate Professor at Boston University in the Department of Mechanical Engineering, Division of Materials Science and Engineering, and Department of Physics.  He is primarily interested in the mechanical properties of atomically thin materials such as graphene. He received his B.S. degree in Physics from Florida International University (2000) and a Ph.D. in Physics (2008) from Cornell University where he studied the electrical and mechanical properties of graphene.  After finishing his Ph.D., he spent 3 months as a postdoctoral researcher in the Laboratory of Atomic and Solid State Physics at Cornell University studying nanoelectromechanical systems. Before moving to Boston University, he was an Assistant Professor of Mechanical Engineering at University of Colorado at Boulder from 2008-2013
\end{IEEEbiography}

\begin{IEEEbiography}[{\includegraphics[width=1in,height=1.25in,clip,keepaspectratio]{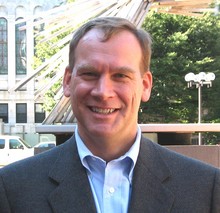}}]{Bennett B. Goldberg} was born in Boston, Mass. in 1959. He received the B.A. degree from Harvard College, MA, in 1982 and the M.S. and Ph.D. degrees in physics from Brown University, Providence, RI, in 1984 and 1987, respectively. Following a Bantrell Post-doctoral appointment at the Massachusetts Institute of Technology and the Francis Bitter National Magnet Lab, he joined the physics faculty at Boston University in 1989. In 2016, Goldberg became the Director of the Searle Center for Advancing Learning and Teaching, the Assistant Provost of Learning and Teaching, and a Professor of Physics and Astronomy at Northwestern University. He combines leadership in local and national projects to support access to and success in higher education by marginalized and traditionally underrepresented students with active research interests in strain physics of 2D crystals; super-resolution and near-field imaging of semiconducting and biological systems; and biosensing and biodetection of single viruses and nanoparticles. 

\end{IEEEbiography}

\begin{IEEEbiography}[{\includegraphics[width=1in,height=1.25in,clip,keepaspectratio]{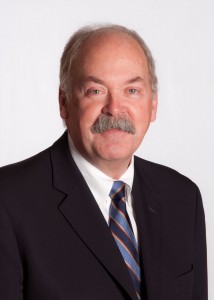}}]{David J. Bishop} (M'11) received the B.S. degree in physics from Syracuse University in 1973, and the M.S. and Ph.D. degrees in physics from Cornell University in 1977 and 1978, respectively. He joined the AT\&T-Bell Laboratories Bell Labs in 1978 as a Post-Doctoral Member of Staff and became a Member of the Technical Staff in 1979. In 1988, he was made a Distinguished Member of the Technical Staff and later he was promoted as the Department Head, Bell Laboratories. He was the President of Government Research and Security Solutions with Bell Labs, Lucent Technologies. He was the Chief Technology Officer and Chief Operating Officer with LGS, the wholly-owned subsidiary of Alcatel-Lucent dedicated to serving the U.S. federal government market with advanced research and development solutions. He is currently the Head of the Division of Materials Science and Engineering, Boston University, a Professor of Physics, and also a Professor of Electrical Engineering. He is a Bell Labs Fellow and in his previous positions with Lucent served as a Nanotechnology Research VP for Bell Labs, Lucent Technologies, and the President of the New Jersey Nanotechnology Consortium and the Physical Sciences Research VP. He is a member and a fellow of the American Physical Society, and a member of the MRS. He was a recipient of the APS Pake Prize.
\end{IEEEbiography}

\begin{IEEEbiography}[{\includegraphics[width=1in,height=1.25in,clip,keepaspectratio]{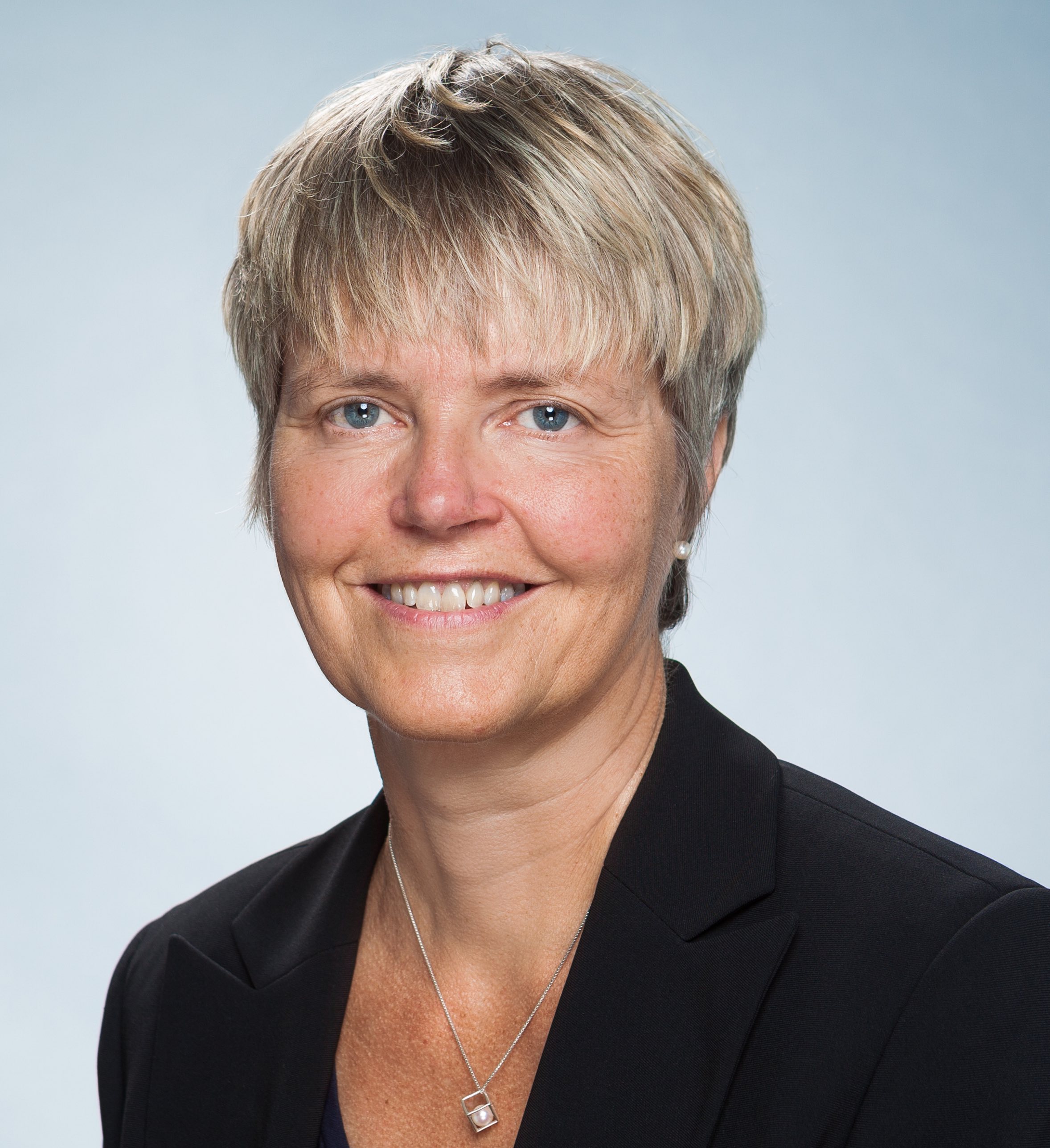}}]{Anna K. Swan} received a BSc and MS degree in Physics Engineering from Chalmers University, Gothenburg, Sweden, and a Ph.D. degree in Physics at Boston University, Boston, MA, in 1994. Her dissertation topic was the spin-ordering on NiO(100) surfaces using metastable He  scattering, for which she received two student awards, the Nottingham prize and the Morton M. Traum Award. She joined the Solid State Division at Oak Ridge National Laboratory as a Wigner Fellow. and later as a staff member. In 2005, she joined the Electrical and Computer Engineering Department at Boston University where she is currently an Associate Professor and Associate chair of graduate studies. Her research interests  clustered around high-spatial resolution spectroscopy. Currently she is working on 2D materials and their responses to strain and charge using photoluminescence  and  micro-Raman spectroscopy.
\end{IEEEbiography}

\newpage

\section{Supplementary Information for Monolayer MoS\texorpdfstring{$_2$}{2} Strained to 1.3\% with a Microelectromechanical System}

\subsection{Tri-Layer Graphene Raman Analysis of Strain} 
As stated in the main text, we find the work of Garza \emph{et al.} \cite{Garza2014} to be a valuable reference for many important concepts regarding the use of MEMS to strain 2D materials. However in their article they primarily rely on optical measurements of their sample shuttle displacement to determine strain, which neglects the possibility of slipping. Garza \emph{et al.} found for their sample, trilayer graphene, that the G phonon peak shifted at a rate of -0.24 cm$^{-1}$/\%. We compare this shift rate with the expected rate given the literature values for the Gr\"uneisen parameter and shear deformation potential for trilayer graphene ($\gamma_{G}$ = 1.89 $\pm$ .02, and $\beta_{G}$ = 0.71 $\pm$ .06)~\cite{Kitt2013a}, the Poisson's ratio, $\nu$, for Graphite (0.165)~\cite{Blakslee1970}, and the fact that under strain the G peak shifts according to the same formula as the $E'$ peak in MoS$_2$,
   \begin{align}
    \omega_{G}^{\pm} & = \omega_{0G}\lb1-\[\gamma_{G}\(1-\nu\)\mp\frac{\beta_{G}}{2}\(1+\nu\)\]\epsilon\rb \label{eq:EstrainS},
  \end{align}
where $\omega_{0G}$ is the unstrained phonon energy (1584.9)~\cite{Garza2014}, and $\epsilon$ is the uniaxial strain. According to the given parameter values and the G phonon energy dependence on strain, it is expected that $\der{\omega_G^+}{\epsilon} \approx -12$ cm$^{-1}$/\% and $\der{\omega_G^-}{\epsilon} \approx -38$ cm$^{-1}$/\%. This shows that the shift rate reported by Garza \emph{et al.} is two orders of magnitude smaller than expected based on the literature. Such a large discrepancy strongly indicates that Garza \emph{et al.} did not achieve the claimed strain. Their maximum observed shift of -3.1 cm$^{-1}$ is equivalent to 0.08\% to 0.26\% strain depending on the G mode.

\subsection{Displacement vs. Power Curve} 
\begin{figure*}[!t]
  \centering
  \includegraphics[width=4in]{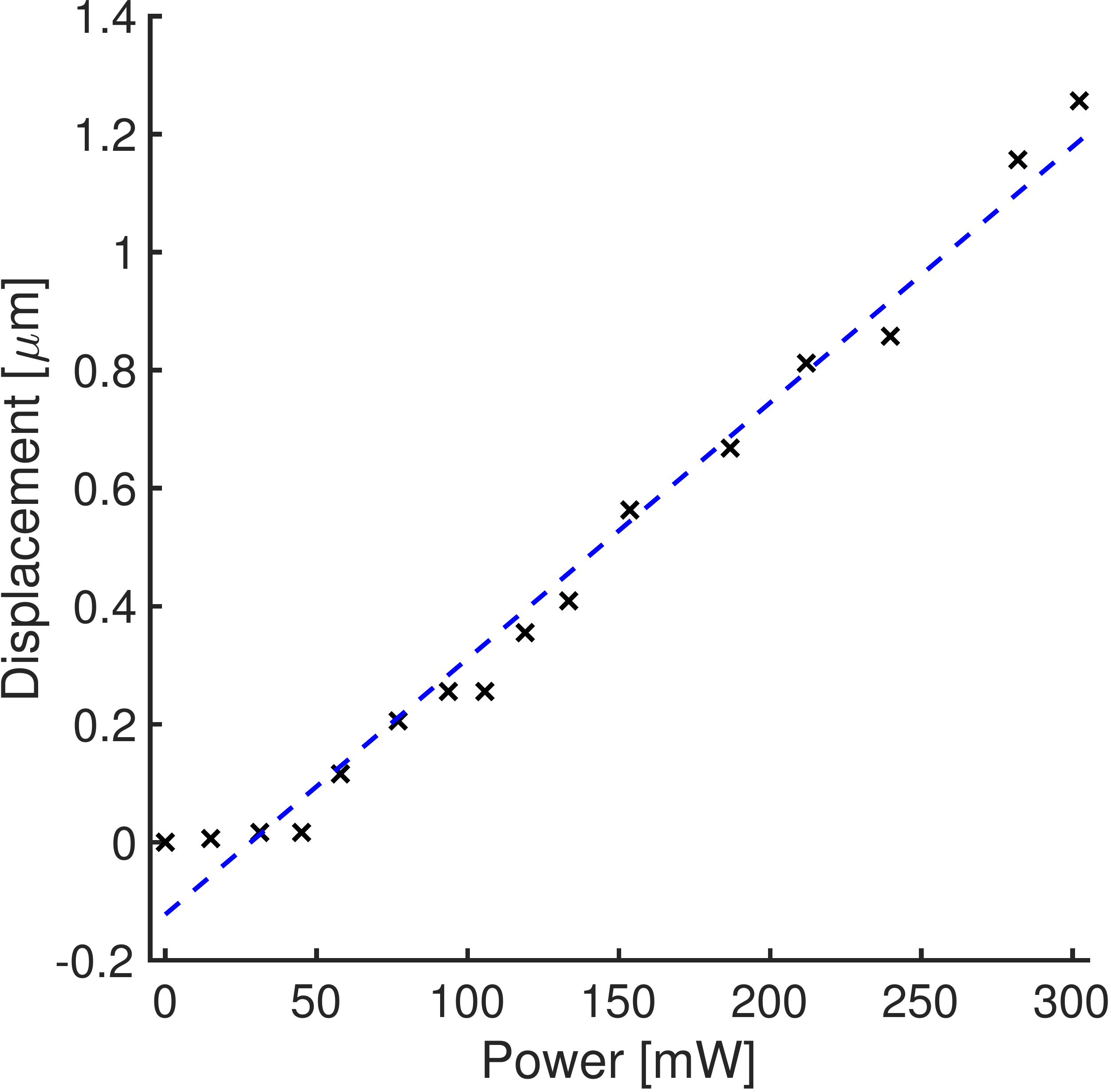}
  \caption{Example shuttle displacement versus power applied curve for one of our thermal actuators.}
  \label{fig:DvsP}
\end{figure*}

Figure~\ref{fig:DvsP} shows an example optical measurement of displacement versus power applied to on of our thermal actuators. The slope is 4.3 nm/mW, which we used to estimate the rate at which we applied strain to our samples. For example if we had a sample with a 3 $\mu$m gap, our nominal gap size, and we wanted to increase the strain in steps of 0.2\%, then we would increase the power in steps of $\approx$1.4 mW.

\subsection{Effective Poisson's Ratio} 
The equations for mechanical equilibrium of a coupled bulk and 2D system are derived below start with the free energy equations for the bulk and 2D system, and couple them using Lagrange multipliers. The total free energy is
\begin{align}
  H = \int_{\Sigma_B} \mspace{-16.0mu} \mathrm{d}V F_B\(\epsilon^B\) + \tau\int_{\Sigma_S} \mspace{-14.0mu} \mathrm{d}S \[F_S\(\epsilon^S\) + \lambda_{ij}\(\eval{\epsilon^B_{ij}}{z=0}{} - \epsilon^S_{ij}\)\]
\end{align}
where $F_B$, $\epsilon^B$ and $F_S$, $\epsilon^S$ are the free energy densities and strain tensors of the bulk material and surface (2D system), $\tau$ is the effective thickness of the 2D material, $\Sigma_B$ is the volume of the bulk material, $\Sigma_S$ is the surface of the 2D system, and $\lambda_{ij}$ are Lagrange multipliers to constrain the strain of the 2D material and the bulk at the boundary, $z=0$, to be equal. Since $\epsilon$ is symmetric, so too must $\lambda$ otherwise there would be too many constraints. Summation over repeated induces is assumed. Note that $\epsilon^B$ is a three dimensional tensor, while $\epsilon^S$ is only a two dimensional tensor and only has support over the domain $z=0$.

Now we can take functional derivatives with respect to the displacement fields $u^B$ and $u^S$, bulk and surface, to determine the equilibrium equations and boundary conditions. We'll also define the stress tensor $\sigma_{ij}^{X}:=\pder{F_X}{\epsilon^X_{ij}}$ where $X$ can be $B$ or $S$.
\begin{align}
\frac{\delta H}{\delta u^B_i} & = \int_{\Sigma_B} \mspace{-16.0mu} \mathrm{d}V \sigma^B_{jk}\frac{1}{2}\(\delta_{ik}\partial_j \delta^{(3)}\(x\) + \delta_{ij}\partial_k \delta^{(3)}\(x\)\) \nonumber \\
& + \tau\int_{\Sigma_S} \mspace{-14.0mu} \mathrm{d}S \lambda_{jk}\frac{1}{2}\[\delta_{ik}\partial_j \delta^{(2)}\(x\) + \delta_{ij}\partial_k \delta^{(2)}\(x\)\]\delta\(z\) \\
& = \int_{\Sigma_B} \mspace{-16.0mu} \mathrm{d}V \[\partial_j\(\sigma^B_{ij} \delta^{(3)}\(x\)\) - \delta^{(3)}\(x\)\partial_j\sigma^B_{ij}\] \nonumber \\
& + \delta\(z\)\tau\int_{\Sigma_S} \mspace{-14.0mu} \mathrm{d}S \[\partial_j\(\lambda_{ij} \delta^{(2)}\(x\)\) - \delta^{(2)}\(x\)\partial_j\lambda_{ij}\] \\
& = -\int_{\Sigma_B} \mspace{-14.0mu} \mathrm{d}V \delta^{(3)}\(x\) \partial_j\sigma^B_{ij} + \oint_{\partial\Sigma_B} \mspace{-20.0mu} \mathrm{d}A_j \sigma^B_{ij} \delta^{(3)}\(x\) \nonumber \\ 
& - \delta\(z\)\tau\int_{\Sigma_S} \mspace{-14.0mu} \mathrm{d}S\, \delta^{(2)}\(x\) \partial_j\lambda_{ij} + \delta\(z\)\tau\oint_{\partial\Sigma_S} \mspace{-14.0mu} \mathrm{d}\ell_j \lambda_{ij} \delta^{(2)}\(x\) \\
& = -\int_{\Sigma_B} \mspace{-14.0mu} \mathrm{d}V \partial_j\sigma^B_{ij}\delta^{(3)}\(x\) + \mspace{-20.0mu} \int\limits_{\partial\Sigma_B,z\neq0} \mspace{-20.0mu} \mathrm{d}A_j \sigma^B_{ij} \delta^{(3)}\(x\) \nonumber \\ 
& + \delta\(z\) \mspace{-40.0mu} \int\limits_{\Sigma_S=\partial\Sigma_B,z=0} \mspace{-40.0mu} \mathrm{d}S\[ \sigma^B_{ij}\hat{A}_j - \tau\partial_j\lambda_{ij}\] \delta^{(2)}\(x\) \nonumber \\
& + \delta\(z\)\tau\oint_{\partial\Sigma_S} \mspace{-14.0mu} \mathrm{d}\ell_j \lambda_{ij} \delta^{(2)}\(x\) = 0,
\end{align}
where $\delta^{\(n\)}$ is the $n$-dimensional Kronecker delta function, and $\partial\Sigma_X$ is used to denote the boundary of $\Sigma_X$. In the first line the partial derivative with respect to $\epsilon^B_{jk}$ is taken, and then the functional derivative of $\epsilon^B_{jk}$. In the second line the sums are collapsed making use of the symmetry of $\epsilon$ and $\lambda$, and the first step of integration by parts, $f\partial_j g = \partial_j\(fg\) - g\partial_j f$, is used. In the third line the generalized Stoke's theorem is used to convert integrals over $\Sigma_B$ and $\Sigma_S$ to their boundaries, $\partial\Sigma_B$ and $\partial\Sigma_S$. The integration measures $\mathrm{d}A_j$ and $\mathrm{d}\ell_j$ are oriented outward of the integration domain. In the fourth line the portion of the integral over $\partial\Sigma_B$ that corresponds with $\Sigma_S$, \emph{i.e.} where $z=0$, is combined with the explicit integral over $\Sigma_S$.
Setting each integrand equal to zero we arrive at the equilibrium conditions and boundary conditions for the bulk material.
\begin{align}
\partial_j\sigma^B_{ij} & =0 \\
\eval{\sigma^B_{ij}\hat{A}_j}{\partial\Sigma_B, z\neq0}{} \mspace{-40.0mu} & =0 \\
\eval{\sigma^B_{ij}\hat{A}_j}{z=0}{}-\tau\partial_j\lambda_{ij} & =0 \\
\eval{\lambda_{ij}\hat{\ell}_j}{\partial\Sigma_S}{} & =0
\end{align} 
Note that $\hat{A}_j$ and $\hat{\ell}_j$ are outward normal unit vectors of their domains. The first equation is the standard equilibrium equation for a bulk elastic material, and the second equation is the standard boundary condition for no external forces acting on the material. The third equation gives a rule for balancing the forces between the bulk material and the 2D system, and the fourth provides a no force boundary condition on the 2D system. 

Now we determine the equilibrium equations and boundary conditions of the 2D system by taking the functional derivative with respect the the 2D system's displacement field.
\begin{align}
\frac{\delta H}{\delta u^S_i} & = \tau\int_{\Sigma_S} \mspace{-14.0mu}\mathrm{d}S \(\sigma^S_{ij} - \lambda_{ij}\)\partial_j \delta^{(2)}\(x\) \\
& = \tau\int_{\Sigma_S}\mspace{-14.0mu} \mathrm{d}S \lb\,\partial_j\[\(\sigma^S_{ij} - \lambda_{ij}\) \delta^{(2)}\(x\)\] - \delta^{(2)}\(x\)\partial_j\(\sigma^S_{ij} - \lambda_{ij}\)\rb \\
& = \tau \lb\oint_{\partial\Sigma_S} \mspace{-14.0mu}\mathrm{d}\ell_j\(\sigma^S_{ij} - \lambda_{ij}\) \delta^{(2)}\(x\) - \int_{\Sigma_S}\mspace{-14.0mu}\mathrm{d}S\[\partial_j\(\sigma^S_{ij} - \lambda_{ij}\) \delta^{(2)}\(x\)\]\rb = 0
\end{align}
This derivation followed the same exact steps as for $\frac{\delta H}{\delta u^B_i}$, but there are fewer terms and there is no complication of the boundary of one domain being part of another integrals domain. Setting each integrand equal to zero we have
\begin{align}
\eval{\(\sigma^S_{ij} - \lambda_{ij}\)\hat{\ell}_j}{\partial\Sigma_S}{} & =0 \\
\partial_j\(\sigma^S_{ij} - \lambda_{ij}\) & =0.
\end{align} 

We can now easily eliminate the Lagrange multipliers yielding the following set of equations
\begin{align}
0 & = \partial_j\sigma^B_{ij},\\
0 & = \eval{\sigma^B_{ij}\hat{A}_j}{\partial\Sigma_B, z\neq0}{}, \\
0 & = \eval{\sigma^B_{ij}\hat{A}_j}{z=0}{}-\tau\partial_j\sigma^S_{ij}, \label{eq:ElasticCoupling}\\
0 & = \eval{\sigma^S_{ij}\hat{\ell}_j}{\partial\Sigma_S}{} \\
0 & = \eval{\epsilon^B_{ij}}{z=0}{} - \epsilon^S_{ij}. \label{eq:StrainCoupling}
\end{align}
As mentioned earlier the first two equations are the standard equilibrium equation and zero force boundary conditions of a bulk material. The third term relates the boundary force of the bulk to the interior force on the 2D material. Note that when $i=z$, the second term is zero. Further, for a system where the 2D material is at the $z=0$ plane, $\hat{A}_j=\hat{z}$, so this equation reads $0 = \eval{\sigma^B_{iz}}{z=0}{}-\tau\partial_j\sigma^S_{ij}$. The fourth equation is the standard no force boundary condition, but this time for a 2D system. The final equation is a reminder that though we have eliminated the Langrange multipliers, we still need to adhere to the constraints they impose. This is very important because otherwise you might think that you could decouple the bulk from the 2D system by creating a uniform strain distribution with zero off diagonal strain components, which would trivially satisfy Equation~\ref{eq:ElasticCoupling}.

We'd really like to solve these equations in the situation of uniaxial strain in isotropic bulk and 2D materials. In the bulk situation this is solved by hypothesizing a uniform strain distribution, which trivially satisfies the equilibrium equation, and then solving for the strain components that satisfy the no force boundary conditions. If we tried to use the same method to solve these coupled equations it isn't hard to see that we will fail to satisfy the requirement that strains between the bulk and 2D system are equal. In short we'd find
\begin{align}
\epsilon^B_{xy} = \epsilon^B_{xz} = \epsilon^B_{yz} = 0, \\
\epsilon^B_{xx} = \epsilon, \quad \epsilon^B_{yy} = \epsilon^B_{zz} = -\nu_B \epsilon, \\
\epsilon^S_{xy} = 0, \\
\epsilon^S_{xx} = \epsilon, \quad \epsilon^S_{yy} = -\nu_S \epsilon,
\end{align}
where $\epsilon$ is the stain along the $x$ axis (major strain axis), and $\nu_B$ and $\nu_S$ are the Poisson's ratios of the bulk and 2D system. Unless $\nu_B = \nu_S$ these strains do not satisfy the requirement of Equation~\ref{eq:StrainCoupling}. There simply isn't a uniform solution to the coupled bulk, 2D material equations.

However, we can find a uniform strain field that does not violate the coupled equations so egregiously. Let's begin with what we want to respect most, Equation~\ref{eq:ElasticCoupling}, and hypothesize that $\epsilon^B_{xx}=\epsilon^S_{xx}=\epsilon$ and $\epsilon^B_{yy}=\epsilon^S_{yy}=-\nu_{\text{eff}}\epsilon$. Now we try to satisfy as many boundary conditions as possible. Very easily we'll find $\epsilon^B_{xz}=\epsilon^B_{yz}=\epsilon^B_{xy}=\epsilon^S_{xy}=0$. Setting $\sigma^B_{zz}=0$ we'll find that $\epsilon^B_{zz} = -\nu_B\frac{1-\nu_{\text{eff}}}{1-\nu_B}\epsilon$. The remaining boundary conditions are $\sigma^B_{yy}=0$ and $\sigma^S_{yy}=0$. We could choose $\nu_{\text{eff}}$ to satisfy one of these, but that would mean setting $\nu_{\text{eff}}$ equal to $\nu_B$ or $\nu_S$. Instead we choose to satisfy neither, and let energy minimization select the best $\nu_{\text{eff}}$. Integrating the free energy density over the thickness of the bulk we find
\begin{align}
\int\mathrm{d}zF_B\(\epsilon^B\) + \tau_S F_S\(\epsilon^S\) & = \frac{\tau_B}{2}\(\sigma^B_{xx}\epsilon^B_{xx} + \sigma^B_{yy}\epsilon^B_{yy} + \sigma^B_{zz}\epsilon^B_{zz}\) \nonumber \\
& + \frac{\tau_S}{2}\(\sigma^S_{xx}\epsilon^S_{xx} + \sigma^S_{yy}\epsilon^S_{yy}\) \\
& = \frac{\epsilon^2}{2}\sum_{X\in\lb B,S\rb} \frac{\tau_X E_X}{\(1+\nu_X\)\(1-\nu_X\)}\(1 - 2\nu_X\nu_{\text{eff}} + \nu_{\text{eff}}^2\)
\end{align}
where $\tau_B$ and $\tau_S$ and $E_B$ and $E_S$ are the thicknesses and Young moduli of the bulk and 2D material. Let $\mathcal{K}_X:=\frac{\tau_X E_X}{\(1+\nu_X\)\(1-\nu_X\)}$, then we can simply set the derivative with respect to $\nu_{\text{eff}}$ to zero in order to minimize the energy density. The solution is
\begin{align}
\nu_{\text{eff}} = \frac{\mathcal{K}_B\nu_B + \mathcal{K}_S\nu_S}{\mathcal{K}_B + \mathcal{K}_S}.
\end{align}

This solution has a nice intuitive balance. If the product of the thickness and the Young's modulus of the 2D material is small compared with the substrate, then the effective Poisson's ratio is that of the substrate. This is the typical assumption for 2D materials on bulk substrates. However, when the product of the thickness and the Young's modulus of the 2D material is large compared with the substrate, then the effective Poisson's ratio is that of the 2D material. In particular, if there isn't a substrate, $\mathcal{K}_B = 0$, we recover what we'd expect, $\nu_{\text{eff}}=\nu_S$.


\end{document}